\documentclass[twocolumn,times]{aastex63}
\hypersetup{linkcolor=red,citecolor=blue,filecolor=cyan,urlcolor=magenta}
\usepackage{rotating}
\usepackage{multirow}

\newcommand{\kms}{\mbox{km\,s$^{-1}$}}

\newcommand{\htwo}{\mbox{$\rm H_2$}}

\newcommand{\ceighteeno}{\mbox{$\rm C^{18}O$}}

\newcommand{\gcm}{\mbox{g cm$^{-2}$}} 
\newcommand{\mjypbm}{\mbox{mJy\,beam$^{-1}$}}

\newcommand{\jypbm}{\mbox{Jy\,beam$^{-1}$}}

\newcommand{\msun}{\mbox{\,$M_\odot$}}
\newcommand{\msunyr}{\mbox{\,$M_\odot$ yr$^{-1}$}}

\graphicspath{{./}{figures/}}

\received{}
\revised{}
\accepted{}
\submitjournal{ApJ}

\shorttitle{LZ-STAR. I. Ordered massive star formation in the outer Galaxy}
\shortauthors{Cheng et al.}

\begin{document}

\title{Low-Metallicity Star Formation Survey in Sh2-284 (LZ-STAR).\\ I. Ordered massive star formation in the outer Galaxy}

\correspondingauthor{Yu Cheng}
\email{ycheng.astro@gmail.com}

\author[0000-0002-8691-4588]{Yu Cheng}
\affiliation{National Astronomical Observatory of Japan, 2-21-1 Osawa, Mitaka, Tokyo, 181-8588, Japan}
\affiliation{Dept. of Astronomy, University of Virginia, Charlottesville, Virginia 22904, USA}

\author[0000-0002-3389-9142]{Jonathan C. Tan} 
\affiliation{Dept. of Astronomy, University of Virginia, Charlottesville, Virginia 22904, USA}
\affiliation{Dept. of Space, Earth \& Environment, Chalmers University of Technology, 412 93 Gothenburg, Sweden}

\author[0000-0002-5306-4089]{Morten Andersen}
\affiliation{European Southern Observatory, Karl Schwarzschild Str. 2, 85748, Garching, Germany}

\author[0000-0003-4040-4934]{Rub\'{e}n Fedriani}
\affiliation{Dept. of Space, Earth \& Environment, Chalmers University of Technology, 412 93 Gothenburg, Sweden}
\affiliation{Instituto de Astrof\'{i}sica de Andaluc\'{i}a, CSIC, Glorieta de la Astronomía s/n, 18008 Granada, Spain}

\author[0000-0001-7511-0034]{Yichen Zhang}
\affiliation{Dept. of Astronomy, University of Virginia, Charlottesville, Virginia 22904, USA}
\affiliation{Department of Astronomy, School of Physics and Astronomy, Shanghai Jiao Tong University, 800 Dongchuan Rd., Minhang, Shanghai 200240, China}

\author[0000-0002-9573-3199]{Massimo Robberto}
\affiliation{Space Telescope Science Institute, 3700 San Martin Drive, Baltimore, MD 21218, USA}

\author[0000-0002-7402-6487]{Zhi-Yun Li}
\affiliation{Dept. of Astronomy, University of Virginia, Charlottesville, Virginia 22904, USA}

\author[0000-0002-6907-0926]{Kei E. I. Tanaka}
\affiliation{Department of Earth and Planetary Sciences, Institute of Science Tokyo, Meguro, Tokyo 152-8551, Japan}


\begin{abstract}
Star formation is a fundamental, yet poorly understood, process of the Universe. It is important to study how star formation occurs in different galactic environments. Thus, here, in the first of a series of papers, we introduce the Low-Metallicity Star Formation (LZ-STAR) survey of the Sh2-284 (hereafter S284) region, which, at $Z\sim 0.3-0.5Z_\odot$, is one of the lowest-metallicity star-forming regions of our Galaxy. LZ-STAR is a multi-facility survey, including observations with {\it JWST}, {\it ALMA}, {\it HST}, {\it Chandra} and {\it Gemini}. As a starting point, we report {\it JWST} and {\it ALMA} observations of one of the most massive protostars in the region, S284p1. The observations of shock-excited molecular hydrogen reveal a symmetric, bipolar outflow originating from the protostar, spanning several parsecs, and fully covered by the {\it JWST} field of view and the {\it ALMA} observations of CO(2-1) emission. This allows us to infer that the protostar has maintained a relatively stable orientation of disk accretion over its formation history. The {\it JWST} near-IR continuum observations detect a centrally illuminated bipolar outflow cavity around the protostar, as well as a surrounding cluster of low-mass young stars. We develop new radiative transfer models of massive protostars designed for the low metallicity of S284. Fitting these models to the protostar's spectral energy distribution implies a current protostellar mass of $\sim10\:M_\odot$ has formed from an initially $\sim100\:M_\odot$ core over the last $\sim3\times10^5$ years. Overall, these results indicate that massive stars can form in an ordered manner in low-metallicity, protocluster environments.
\end{abstract}

\keywords{ISM: clouds --- stars: formation}

\section{Introduction}\label{sec:intro}

Star formation is a fundamental process for the cosmos with many open questions. To gain a comprehensive understanding, it is crucial to study star-forming clouds in a wide range of environments. Among various environmental factors, metallicity ($Z$) stands out as a particularly interesting variable. Since cosmic metallicity increases in time with the evolution of universe, the formation of stars at low metallicities, especially the determination of their characteristic mass scales, is important for understanding properties of high redshift galaxies and the early history of the Milky Way. Metallicity may affect star formation via influencing heating and cooling processes and the ionization fraction in molecular gas, which may then regulate collapse, fragmentation, and stellar feedback. For example, theoretical and observational studies indicate that metallicity may affect global star formation properties, including the stellar initial mass function (IMF) \citep[e.g.,][]{Marks12,Tanaka18,Li23}. The lifetimes of protoplanetary disks may also be impacted by the metallicity of the environment \citep{2024ApJ...977..214D}.

Due to limitations in sensitivity and spatial resolution, observational studies capable of spatially resolving  low metallicity star-forming cores and the resulting young stellar populations have primarily focused on relatively nearby targets, such as the Large and Small Magellanic Clouds (LMC and SMC) and the extreme outer Galaxy (defined as regions with galactocentric distances greater than 18~kpc; \citealt{Yasui06}). These regions, with sub-solar metallicities of approximately $\sim 0.2-0.5Z_\odot$, offer low-metallicity environments that may resemble those of high-redshift galaxies or the early Milky Way. With advancements in recent observational facilities, significant progress has been made. The unprecedented capabilities of JWST, for example, have enabled the detection of dusty, sub-solar mass young stellar objects (YSOs) in the NGC~346 region of the SMC \citep{Jones23}. \citet{Yasui24} reported detection of brown dwarfs and a mass function with a relatively low peak mass in the outer Galaxy region, Digel~Cloud~2. For embedded YSOs still in their early stages of evolution, {observations with interferometers such as ALMA have significantly advanced our understanding of the physical processes involved during massive star formation}, including characterization of filaments \citep{Tokuda23,Tokuda25}, chemically rich hot cores \citep{Shimonishi16, Shimonishi20, Shimonishi21, Shimonishi23}, molecular outflows \citep{Fukui15, Tokuda22}, and jets with associated rotating toroids \citep{McLeod18, McLeod24}. These studies provide valuable insights into star formation in low-metallicity environments. However, a systematic investigation that combines the characterization of star-forming gas and the newly-formed stellar content at low metallicity remains lacking. 

\subsection{LZ-STAR Survey Overview}

\begin{figure*}[ht!]
\centering
\includegraphics[width=1.\textwidth]{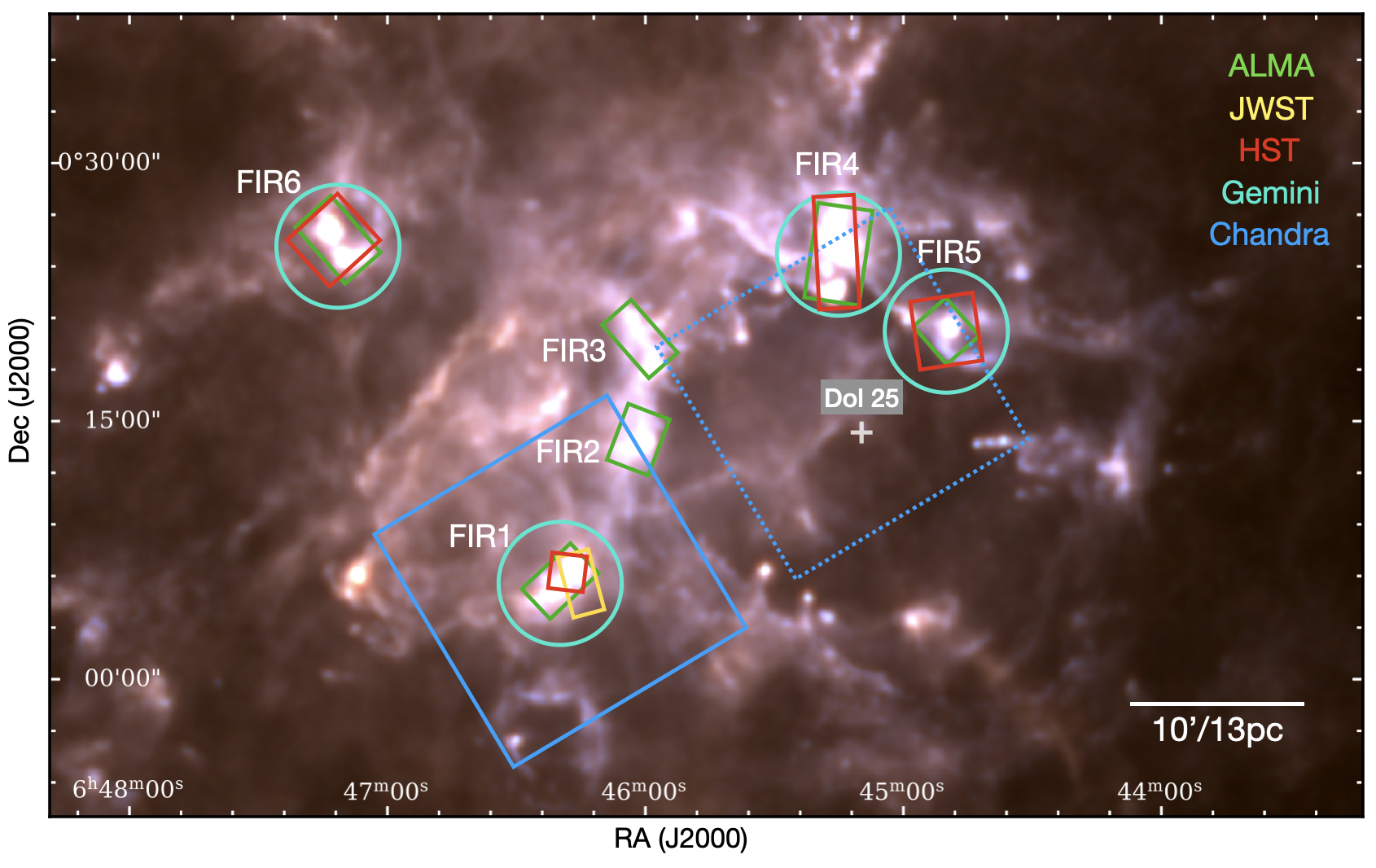}
\caption{Color composite image of the Sh2-284 (S284) region constructed by combining Herschel 250~$\mu$m (blue), 350~$\mu$m (green), and 500~$\mu$m (red) images. 
The colored rectangles or circles indicate the FOV of our observations from different facilities including ALMA, JWST, HST, Gemini, and Chandra. For Chandra the FOV of an archival observation program targeting the Dolidze~25 cluster is also shown (in dashed rectangle) \citep{Guarcello21}.
}\label{fig:survey}
\end{figure*} 

To obtain a more comprehensive understanding of star formation in low-metallicity conditions, we have undertaken the {\it Low-$Z$ Star Formation Survey of Sh2-284} (LZ-STAR Survey) (Co-PIs: Y. Cheng; J. C. Tan).
Initially identified as a diffuse nebula in the Sharpless catalog \citep{Sharpless59}, Sh2-284 (hereafter S284) is located in the outer Galaxy, approximately 4.5~kpc from the Sun \citep{Negueruela15}. Spectroscopic observations of OB stars within S284 indicate a low metallicity, with abundances of 0.3~dex (Si) and 0.5~dex (O) below solar values \citep{Negueruela15}, i.e., corresponding to metallicities of approximately 1/3–1/2 solar, which are comparable to those of the LMC. As such, S284 is one of the most metal-poor star-forming regions currently known in the Milky Way. In addition, {S284 is a large, active star-forming complex spanning at least 50 pc as projected on the sky}. One of its main features is an HII region ionized by the young open cluster Dolidze~25. Surrounding the HII region are dense molecular clouds with ongoing active star formation \citep[][]{Puga09}. 

\autoref{fig:survey} shows a {\it Herschel} FIR image of the S284 region, along with annotations of the various sub-regions that have been targeted in the LZ-STAR Survey.
The survey involves multiple facilities operating at different wavelengths to trace the star formation at different stages from the dense gas to already formed young stellar clusters. Here we give a brief overview of these observations, with most of the data and results to be presented in future papers in this series. We have identified six regions (S284-FIR1 to FIR6) in S284 with active star formation, as evidenced by the strong dust emission seen in the {\it Herschel} images (\autoref{fig:survey}) and large numbers of young stars in previous infrared surveys \citep{Puga09,Guarcello21}.  

Atacama Large Millimeter/sub-millimeter Array (ALMA) observations were carried out towards S284-FIR1 to FIR6 in Cycles 8 and 11 in band~6 (1.3~mm), achieving a spatial resolution of 0\farcs{5}--1.5\arcsec. As described in more detail in \autoref{sec:obs}, these observations are designed to detect millimeter continuum dust emission from dense cores, as well as molecular line emission from various species tracing both dense gas and protostellar outflows. Additional data from APEX-nFlash230 and SEPIA345, and IRAM-30m-NIKA2 have also been obtained for one of the regions (S284-FIR1, see \autoref{fig:survey}) to provide dust continuum and gas kinematics on large scales.

To study the young stellar population, we have carried out near infrared (NIR) observations with the Gemini-South, the Hubble Space Telescope (HST) and James Webb Space Telescope (JWST). The JWST-NIRCam observations have been performed for one region (S284-FIR1), described in more detail in \autoref{sec:obs}, using six filters from 1.6 to 4.7~$\rm \mu m$ designed to detect continuum emission, as well as Br-$\alpha$ and excited $\rm H_2$ emission. HST-WFC3/IR observations in F125W/F139M/F160W filters (Cycle 21), cover four of the six regions (FIR1, FIR4, FIR5, FIR6) with the primary goal of characterizing YSO properties. 
Similarly, the Gemini-South observations with Flamingos-2 in J, H and Ks filters also cover these four regions and are also sensitive to YSOs. In addition, Chandra ACIS-I observations (Cycle 23, 270~ks) have also been obtained to facilitate identification of pre-main sequence stars in a 17\arcmin$\times$17\arcmin{} FOV centered at S284-FIR1.  

In this paper, the first of a series presenting results from the LZ-STAR Survey, we present ALMA/band~6 and JWST/NIRCam observations towards the dense clump S284-FIR1 and report a candidate massive protostellar object with its outflow structure well resolved in both CO and the \htwo{} 4.7~$\mu$m line. 
The structure of this paper is as follows: the observations are described in \autoref{sec:obs}; the results are presented in \autoref{sec:results}; the implications of these results are discussed in \autoref{sec:diss}; and a summary is given in \autoref{sec:sum}.

\section{Observations}\label{sec:obs}

\subsection{ALMA observations}

The ALMA observations of S284-FIR1 were conducted in band~6 in Cycle 8 (Project ID 2021.1.01706.S, PI: Y. Cheng), during a period from November 2021 to May 2022. We divided the entire survey field (5.5\arcmin$\times$3\arcmin) into three strips, each about 3\arcmin\ long and 1.8\arcmin\ wide. We employed the compact configuration C43-3 to recover scales between 0\farcs{6} and 7\arcsec. Additionally, a 7-m array mosaic was performed for each strip, probing scales up to 29\arcsec. Total power observations of the region were also carried out. We set the central frequency of the correlator sidebands to be the rest frequency of the $\rm{N_2D}^+$(3-2) line {at 231.32~GHz} for SPW0 with a velocity resolution of 0.046~\kms. The second baseband SPW1 was set to $231.00\:$ GHz, i.e., 1.30~mm, to observe the continuum with a total bandwidth of $1.875\:$GHz, which also covers CO(2-1) with a velocity resolution of 1.46~\kms. SPW2 was split to cover $\rm{^{13}CO}$(2-1) and $\rm{C^{18}O}$(2-1) line, both with a velocity resolution of 0.096~\kms. The frequency coverage for SPW3 ranged from 215.85 to $217.54\:$GHz to observe DCN(3-2), DCO$^+$(3-2), SiO(5-4) and $\rm{CH_3OH}(5_{1,4}-4_{2,2})$. 

The raw data were calibrated with the data reduction pipeline using {\it{CASA}} 6.2.1. The continuum visibility data were constructed with all line-free channels. We performed imaging with the {\it tclean} task in {\it Casa} and during cleaning we combined data for all three strips to generate a final mosaic map. The combined 12-m array and 7-m array data were imaged using a Briggs weighting scheme with a robust parameter of 0.5,
which yields a resolution of 0\farcs{64}$\times$0\farcs{54} for continuum. The $1\sigma$ noise levels in the continuum image are 0.11 \mjypbm. We feathered the line images with the TP images to correct for the missing large-scale structures. In this paper we mainly focus on the CO(2-1) line, which has a spatial resolution of 0\farcs{74}$\times$0\farcs{64} and a sensitivity of 3.5~\mjypbm{} per 1.5~\kms{} channel.

\subsection{JWST observations}

JWST observations of S284-FIR1 were conducted on 19th Oct. 2022 with the Near-Infrared Camera \citep[NIRCam;][]{Rieke05,Beichman12} using the F162M, F182M, F200W, F356W, F405N and F470N filters (GO:2318, PI: Y. Cheng, available at MAST: \dataset[doi: 10.17909/5ne9-9e82]{\doi{10.17909/5ne9-9e82}}). The brightest part of FIR1 was centered in one of the CCDs of NIRCam and, to ease scheduling, the orientation was left unconstrained. This ultimately yielded partial overlap of the ALMA FOV (see \autoref{fig:survey}). The exposure time for the six filters were 1739s, 1739s, 644s, 644s, 1739s, and 1739s, respectively. The observations were performed adopting a Fullbox3 tight dither pattern as well as 2 subpixel dithers, to cover a field of approximately 5\arcmin$\times$ 2\arcmin. The 6.5~m diameter of JWST gives an angular resolution of $\sim$ 0\farcs{07} -- 0\farcs{17} over the range of NIRCam wavelengths in this study. The NIRCAM images were re-reduced using the JWST Calibration Pipeline Version=1.12.5 \citep{jwst}. The three SW filters were reduced using a single pixel scale of 0\farcs{03} and the first read was used for increased dynamic range. 
The LW module was reduced in a similar manner, but with a pixel scale of 0\farcs{06}.

\section{Results}\label{sec:results}

\subsection{Star formation in S284-FIR1}

S284-FIR1 is one of the most active star-forming regions in S284. It contains a parsec-scale main clump and an extension of dense gas to the southeast. From a 2D gaussian fit to the {\it Herschel} Hi-GAL FIR dust emission derived column density map \citep{Molinari10}, 
we estimate the clump has a FWHM of about 30\arcsec{} (0.65~pc) and a dust mass of 2.7~\msun{}. 
To estimate a total gas mass, we adopt a metallicity of $-0.3$~dex \citep{Negueruela15}, i.e., $0.5Z_\odot$. Scaling the local Galactic refractory dust-to-gas mass value of 141 \citep{Draine11} for this metallicity yields a ratio of 282 and a total gas mass of about 760~\msun{}. Thus the mass surface density of the clump averaged inside the half-mass radius is $\Sigma_{\rm cl}=0.15\:{\rm g\:cm}^{-2}$. Such clump mass surface densities are known to be sufficient to form massive stars in more typical, higher-metallicity regions of the Galaxy \citep{DeBuizer17,Liu19,2020ApJ...904...75L,Fedriani23,Telkamp24}.

\subsubsection{ALMA 1.3~mm continuum}

\begin{figure*}[ht!]
\centering
\includegraphics[width=1.\textwidth]{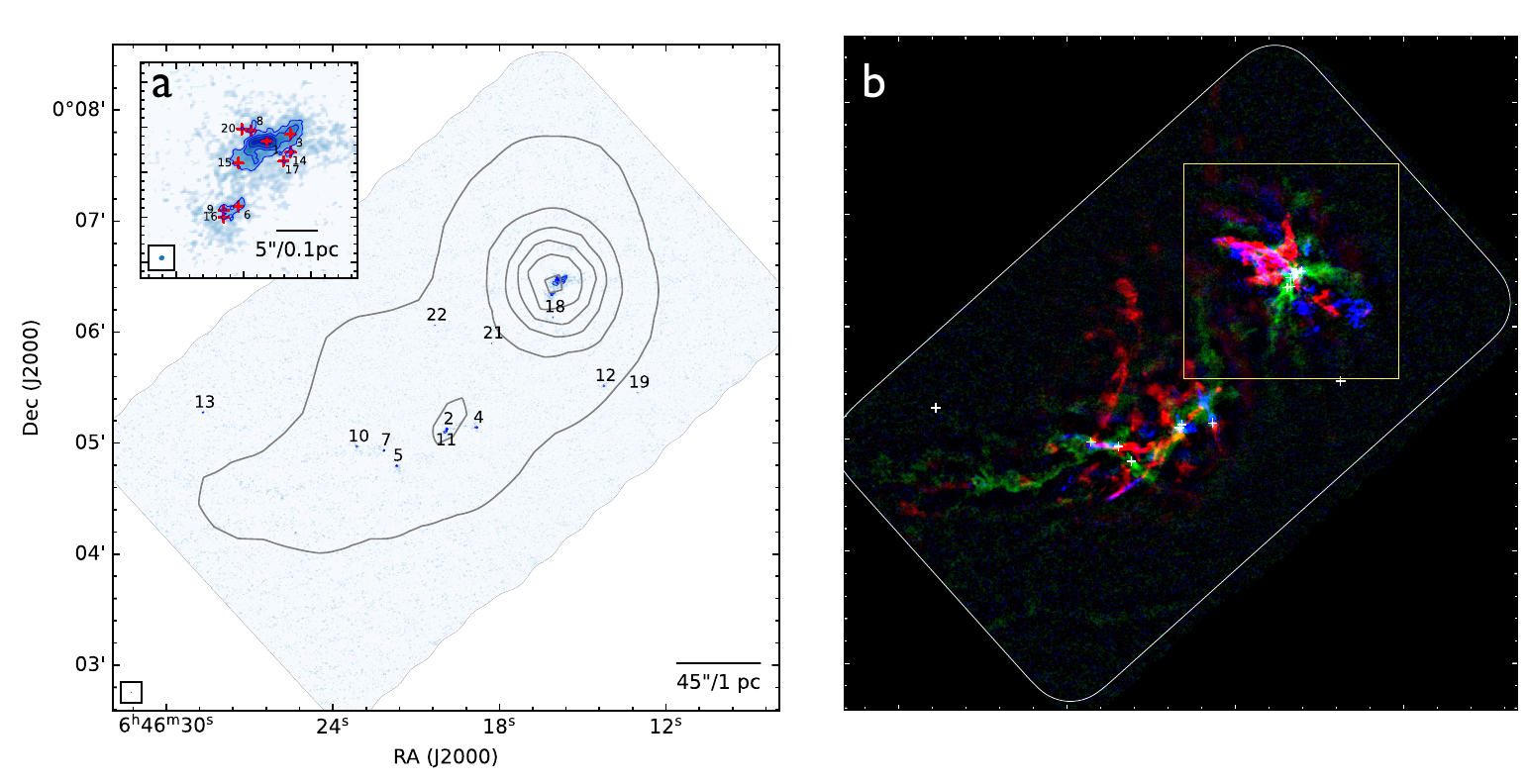}
\caption{{\it (a)} ALMA band~6 (1.3~mm) continuum of the S284-FIR1 region shown in blue color-scale and contours. The contour levels are 0.11~\mjypbm$\times$(5, 10, 20, 40, 80). We overlay the {\it Herschel} 500~$\mu$m image in grey contour for comparison. The identified cores are numbered. A zoom-in view is shown in the top left corner, where the red crosses denote the positions of identified dense cores.  {\it (b)} Three-color image made with ALMA data in the same FOV as {\it (a)}. (green: \ceighteeno{}(2-1) emission integrated over $-$5 to 5~\kms{} relative to the systemic velocity, blue/red: CO(2-1) integrated from 5 to 10~\kms{} blue/redshifted relative to the systemic velocity). {The yellow box indicates position of the main outflow discussed in this work, which is futher shown in \autoref{fig:overview}.  }
}\label{fig:alma}
\end{figure*} 

\autoref{fig:alma} presents the ALMA 1.3~mm dust continuum map of S284-FIR1 with a resolution of 0\farcs{64}$\times$0\farcs{54} (2870~au$\times$2430~au). The image reveals compact sources that are sparsely distributed over the field of view. We identify cores using the {\it astrodendro} algorithm \citep{Rosolowsky08} using parameters of a $4\sigma$ threshold, $1\sigma$ minimum step, and minimum core area of 0.5 beam areas. {These choices represent a reasonable compromise between completeness and the robustness of source identification \citep[e.g.,][]{Cheng18}; this set of parameters are also adopted in a series of core mass function (CMF) studies of \citet{Cheng18}, \citet{2018ApJ...862..105L}, \citet{2021ApJ...916...45O} and \citet{Kinman25}.} We note that cores are identified in the ALMA images before any primary beam correction is applied, which allows a uniform noise level. Then final flux and mass estimates are adjusted to account for the primary beam correction, which is relevant near the edge of the mosaic regions.


The cores are labeled in \autoref{fig:alma}a.
We estimate core gas masses by assuming their emission arises from optically thin isothermal dust, i.e.,
\begin{equation}
    M_{\rm gas} = R_{\rm gd}\frac{d^2F_\nu}{\kappa_\nu B_\nu (T_{d})},
\end{equation}
where $d$ is the distance, $R_{\rm gd}$ is the gas-to-dust mass ratio, $F_\nu$ is the observed flux density, $B_\nu$ is the Planck function, $T_{d}$ is the dust temperature and $\kappa_\nu$ is the dust opacity at the observed wavelength. Following \citet{Cheng18}, we assume a uniform dust temperature of 20~K and adopt $\kappa_{\rm 1.3mm}$ = 0.899~$\rm cm^2\:g^{-1}$ \citep[i.e., from the moderately coagulated thin ice mantle model of][]{Ossenkopf94}. We also adopt $R_{\rm gd}=282$, as discussed above. 

\begin{figure}[ht!]
\epsscale{1.0}\plotone{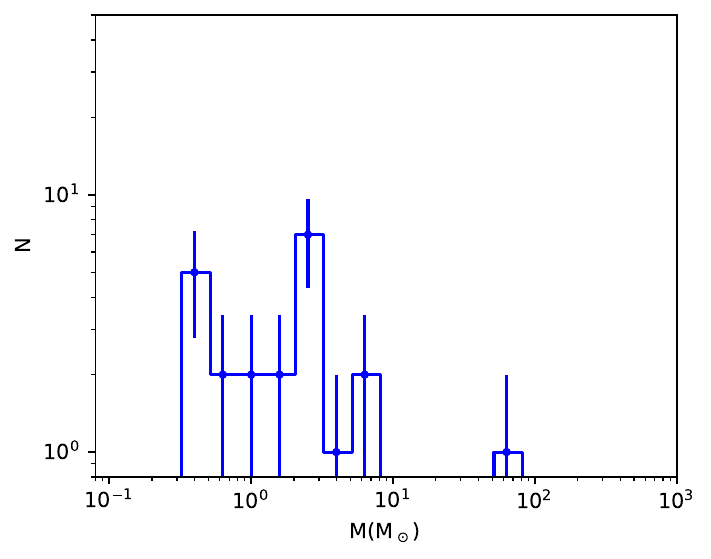}
\caption{The mass distribution of cores detected in the S284-FIR1 region.}
\label{fig:cmf}
\end{figure}

With the above assumptions, the identified cores have masses ranging from 0.37~\msun{} to 68.2~\msun{}. If temperatures of 10~K or 30~K were to be adopted, then the mass estimates would differ by factors of 1.85 and 0.677, respectively. Thus we see that relatively massive cores are present that have the potential to form massive stars. In \autoref{fig:cmf}, we present the CMF derived from the observed core sample. Due to the limited number of cores, the functional form of the distribution is not well constrained. We expect that incorporating additional ALMA observations from other regions in S284 in future work will improve these constraints. {Also note that the flux densities and areas returned by the dendrogram may be underestimated due to isophotal cutoffs, and require further correction before being used for assessing the cumulative properties of the core sample \citep[e.g.][]{Kinman25}.}

From examination of the ALMA data tracing high velocity (both blue and redshifted) CO(2-1) emission (\autoref{fig:alma}b), we see that most of these cores are associated with molecular outflows, often exhibiting highly collimated morphologies. This confirms the presence of active and widespread star formation in S284-FIR1. 
Among the detected outflows, the most prominent one appears to be driven by a source near the center of the main clump in S284-FIR1 { (shown in the yellow box in \autoref{fig:alma})}. It extends several parsecs along a NE-SW axis.
This outflow is mostly likely associated with a massive protostar, which we refer to as S284p1.

\subsubsection{JWST images}

\begin{figure*}[ht!]
\centering
\includegraphics[width=1.\textwidth]{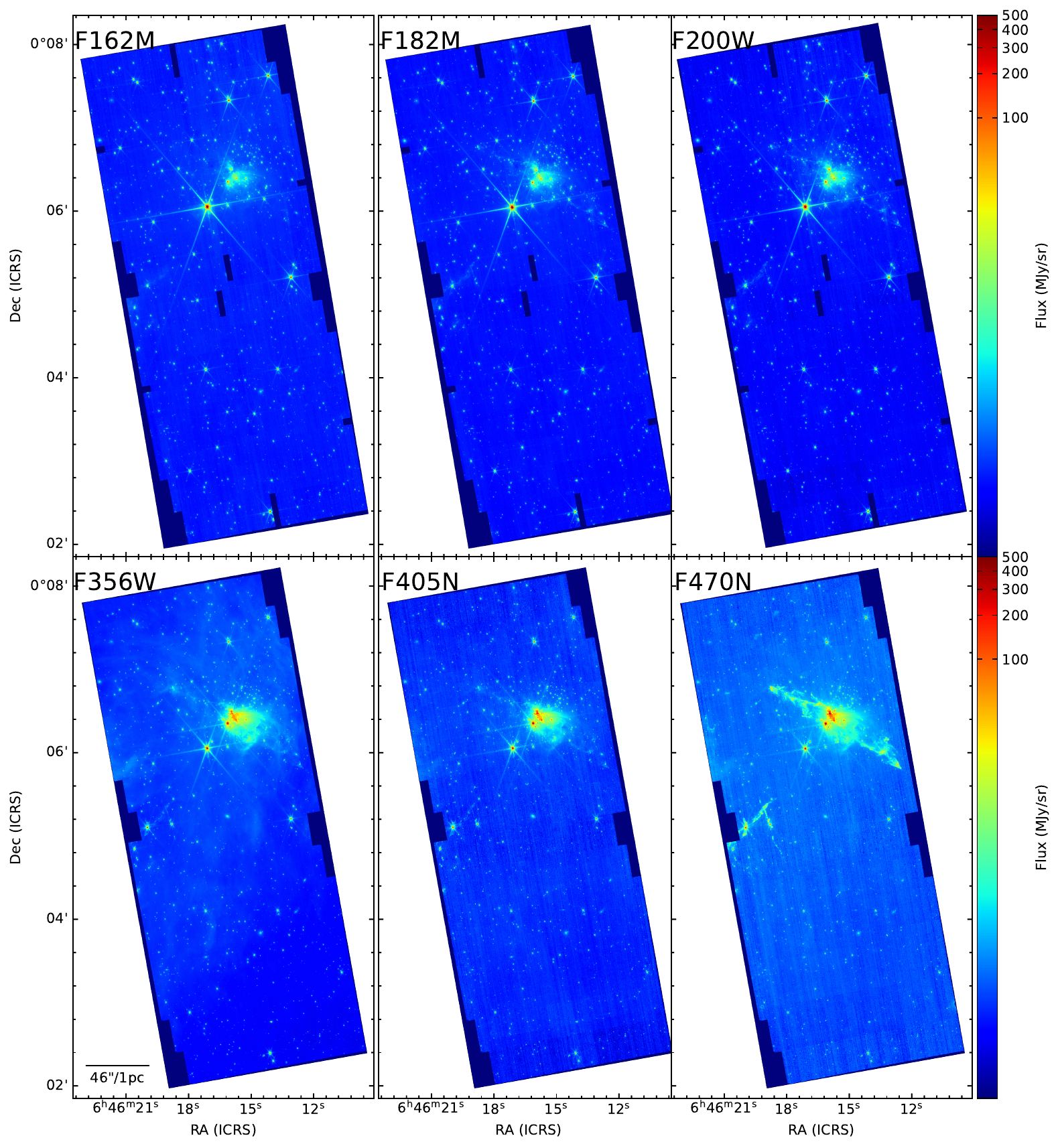}
\caption{JWST images in six filters (F162M, F182M, F200W, F356W, F405N, F470N) towards S284-FIR1 region.}\label{fig:jwst}
\end{figure*} 

In \autoref{fig:jwst}, we present JWST images in all six observed filters toward S284-FIR1. The images reveal a cluster of approximately 200 stars in the northern part of the FOV, coinciding with the main clump in S284-FIR1. Analysis of the colors and fluxes of these stars indicates that they are embedded, recently formed YSOs (Andersen et al., in prep.), with a significant fraction exhibiting infrared excess (Brizawasi et al., in prep.). A prominent IR continuum reflection nebula is visible, particularly at longer wavelengths (F356W, F405N, F470N), powered by sources near the cluster center and extending to the southwest. Additionally, the F470N image, which traces the \htwo{} 4.7~$\mu$m line, reveals a large bipolar jet ejected from the cluster center. This emission line corresponds to the pure rotational transition \htwo{}(0-0~S(9)), with an upper state energy level of 10,263~K, making it an excellent tracer of shocks in jets and outflows \citep{Turner77}. Toward the eastern middle of the F470N FOV, at least two additional collimated jets are visible, which are also detected in the CO data. Some jets or jet knots are also seen in other filters, including wide- and medium-band filters, which are likely partly contributed by the molecular hydrogen lines within their frequency ranges.

\subsection{The massive protostar S284p1}

\subsubsection{Location of the protostar and its immediate environment}\label{sec:S284p1_intro}

\begin{figure*}[!ht]
\includegraphics[width=\textwidth]{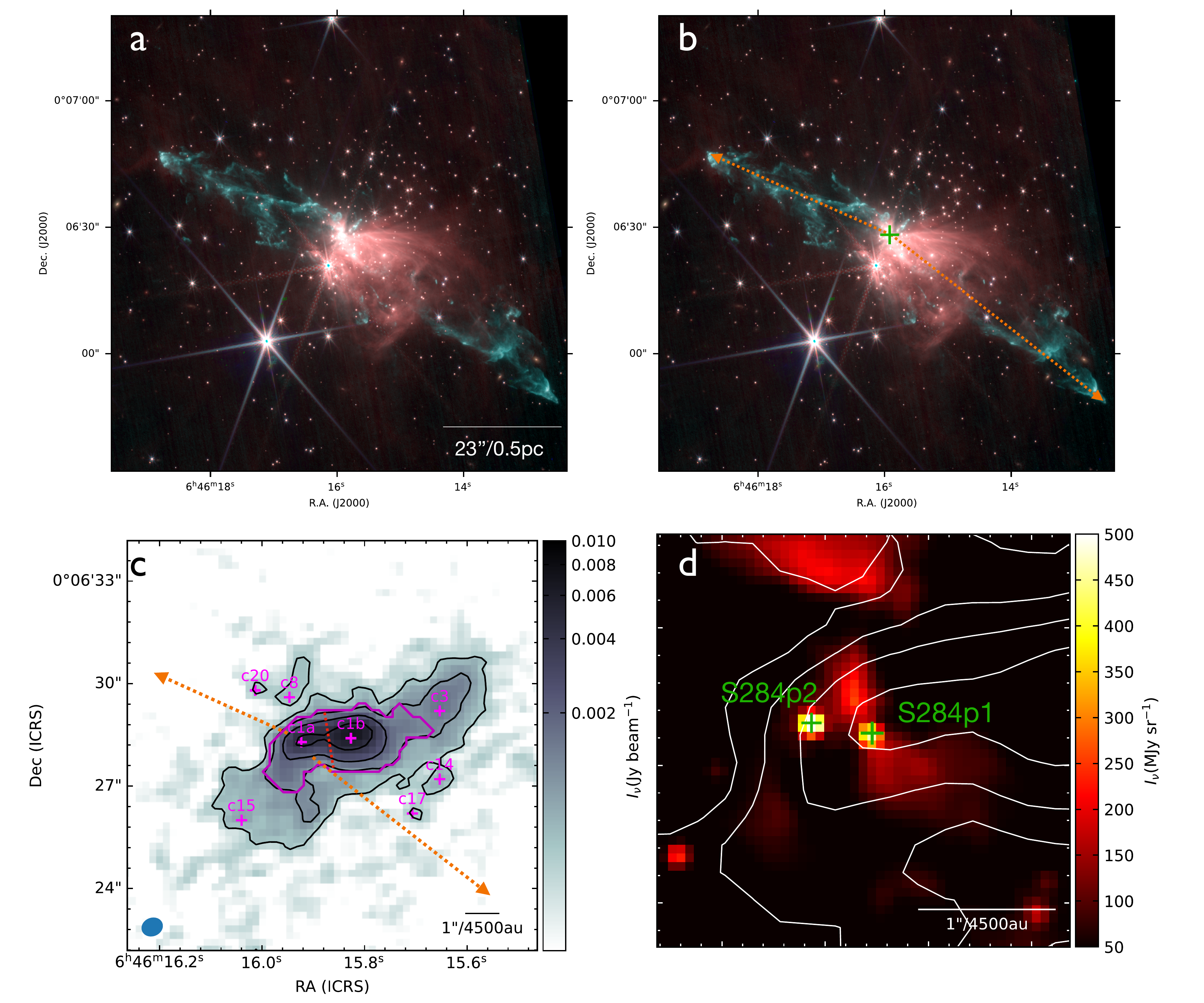}
\caption{Overview of the massive protostar S284p1 and its outflow. {\it (a)} JWST color composite image with $1.62\:{\rm \mu}$m, 2.0$\mu$m, 3.56$\mu$m, 4.70$\mu$m in blue, green, red, and cyan, respectively, and with the latter highlighting excited $\rm H_2$ emission. {\it (b)} Same as {\it (a)}. The green cross indicates the position of S284p1 and the orange dashed arrows show the connection between S284p1 and the outflow termination shocks. {\it (c)} Zoom-in view of S284p1 in 1.3~mm (ALMA) continuum emission (contour levels of 5, 10, 20, 40, 80 $\times$ 0.11~\mjypbm). The magenta crosses indicate the positions of identified dense cores. The dashed arrows indicate the orientation of the outflows as in {\it (b)}. The magenta contour highlights the core boundary for c1 returned by {\it astrodendro}. We manually separate the two peaks (c1a, c1b) in c1, with the boundary shown by the dashed red line. {\it (d)} A further zoom-in view of S284p1, now showing the F356W image. The ALMA 1.3~mm continuum emission is shown with the same contours as in {\it (c)}. 
} 
\label{fig:overview}
\end{figure*}

\begin{figure*}[!ht]
\includegraphics[width=\textwidth]{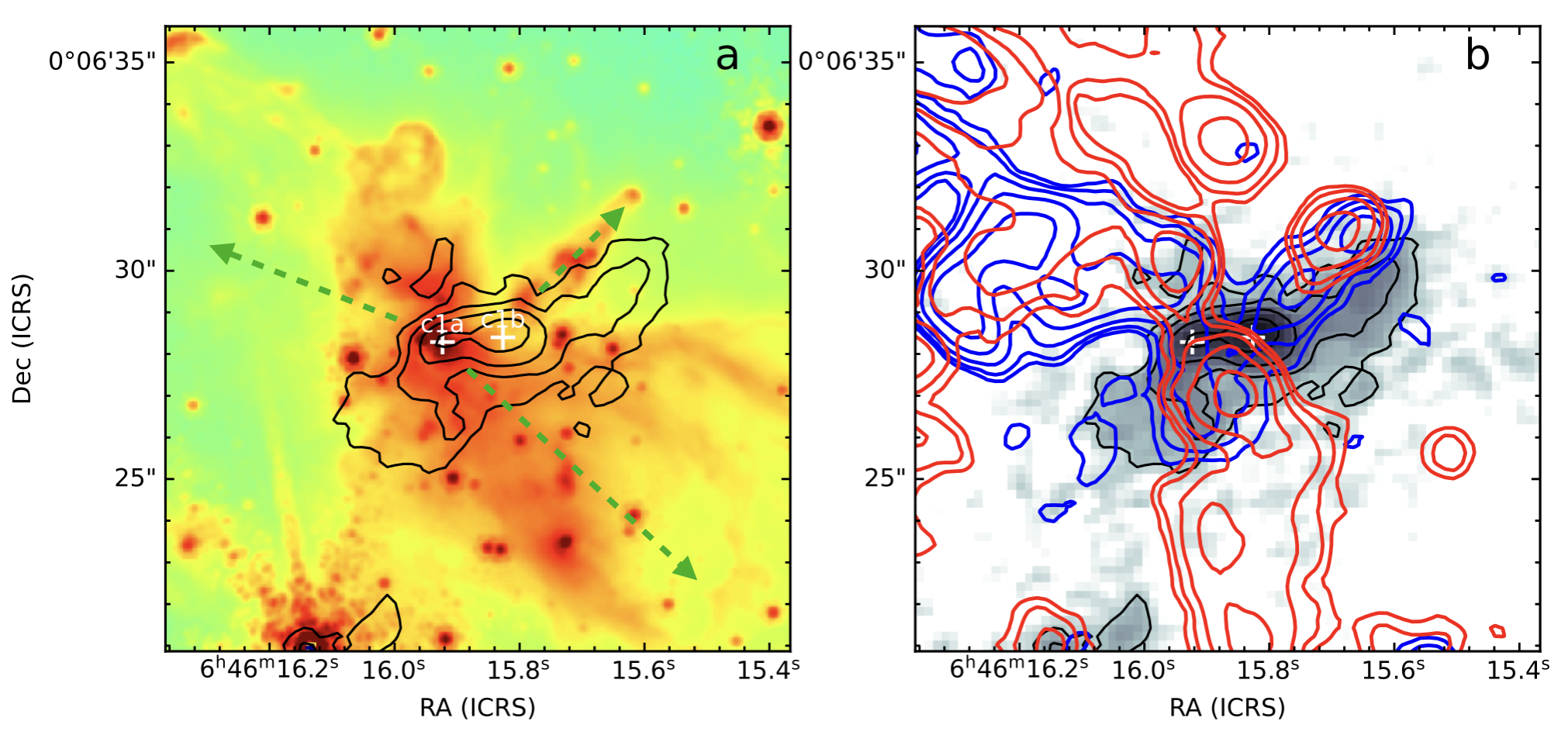}
\caption{Zoom-in to outflow structures from S284p1. {\it (a)} F470N image in colorscale with 1.3~mm continuum overlaid in contours. The white crosses denote the position for the two peaks in core c1. { The dashed green arrows show the orientation of the outflows ejected from c1a and c1b.} {\it (b)} Integrated CO(2-1) high velocity emission in blue and red contours with 1.3~mm continuum overlaid in colorscale and contours. The CO emission is integrated from 5 to 10~\kms{} blue/redshifted relative to the systemic velocity. The contour levels are 0.01~\jypbm\,\kms\,$\times$ (5, 10, 20, 40, 80, 160).
} 
\label{fig:outflow_inner}
\end{figure*} 

A prominent outflow is identified in both CO(2-1) and \htwo{}(0-0~S(9)), which appears to be driven by a source near the center of the main parsec-scale clump in S284-FIR1. This is further illustrated in \autoref{fig:overview}a, where we present a {\it JWST} color composite image combining the F162M, F200W, F356W and F470N bands, represented in red, green, blue and cyan colors, respectively. The F470N (\htwo{} 4.7$\mu$m line) emission highlights a protostellar outflow matching the basic orientation traced by the primary CO(2-1) outflow. The outflow extends about 2\arcmin{} ($\sim$2.6~pc) on the sky and contains numerous knots, arcs, and more diffuse emission features. This is among the longest outflow structures associated with a massive protostar ever observed via molecular hydrogen \citep[e.g.,][]{Bally16}. The large spatial extent and symmetric appearance of the jet suggest that the protostar has maintained a steady orientation throughout the majority of its formation history. The global geometry of the outflow, defined by the directions from the driving source to the termination bow shocks, is highlighted in \autoref{fig:overview}b and discussed in more detail below.

The driving source (S284p1) can be identified from the 1.3~mm continuum map as well as the {\it JWST} images. \autoref{fig:overview}c shows a zoom-in view of the ALMA 1.3~mm continuum emission near the center of S284-FIR1. This image reveals seven dense cores within the clump, among which c1 has the highest millimeter flux densities and an estimated gas mass of 68.2~\msun{} assuming a temperature of 20~K. A close inspection of the image shows that c1 is fragmented into two peaks separated by $\sim$5000~au. These are not identified by {\it astrodendro} as two separate cores since they do not reach the minimum delta value (1$\sigma$) threshold. However, {since the protostar (S284p1, identified from JWST F356W image, as will be discussed below) appears to be located at the position of the eastern peak (c1a)}, we manually divide c1 with the dividing line drawn along the emission minimum ridge. c1a is estimated to have a gas mass of 20.7~\msun, while c1b has a mass of 47.5~\msun. Based on the morphology of the \htwo{} 4.7~$\mu$m and CO(2-1) integrated intensity maps (\autoref{fig:outflow_inner}), we identify c1a as the host for the driving source for the large-scale bipolar outflow. We also note that c1b hosts another protostar that drives another collimated outflow also seen in CO and \htwo{} that is approximately orthogonal to the main outflow from c1a. {This outflow exhibits a single-lobed and collimated appearance, extending to the NW direction from c1b. Given the overlapping blue- and redshifted high velocity CO emission (\autoref{fig:outflow_inner}), we expect this outflow to be oriented close to the plane of sky. Therefore, its small extent likely indicates the protostar hosted by c1b is still in an early evolutionary stage.}

%

\begin{figure*}[!ht]
\includegraphics[width=\textwidth]{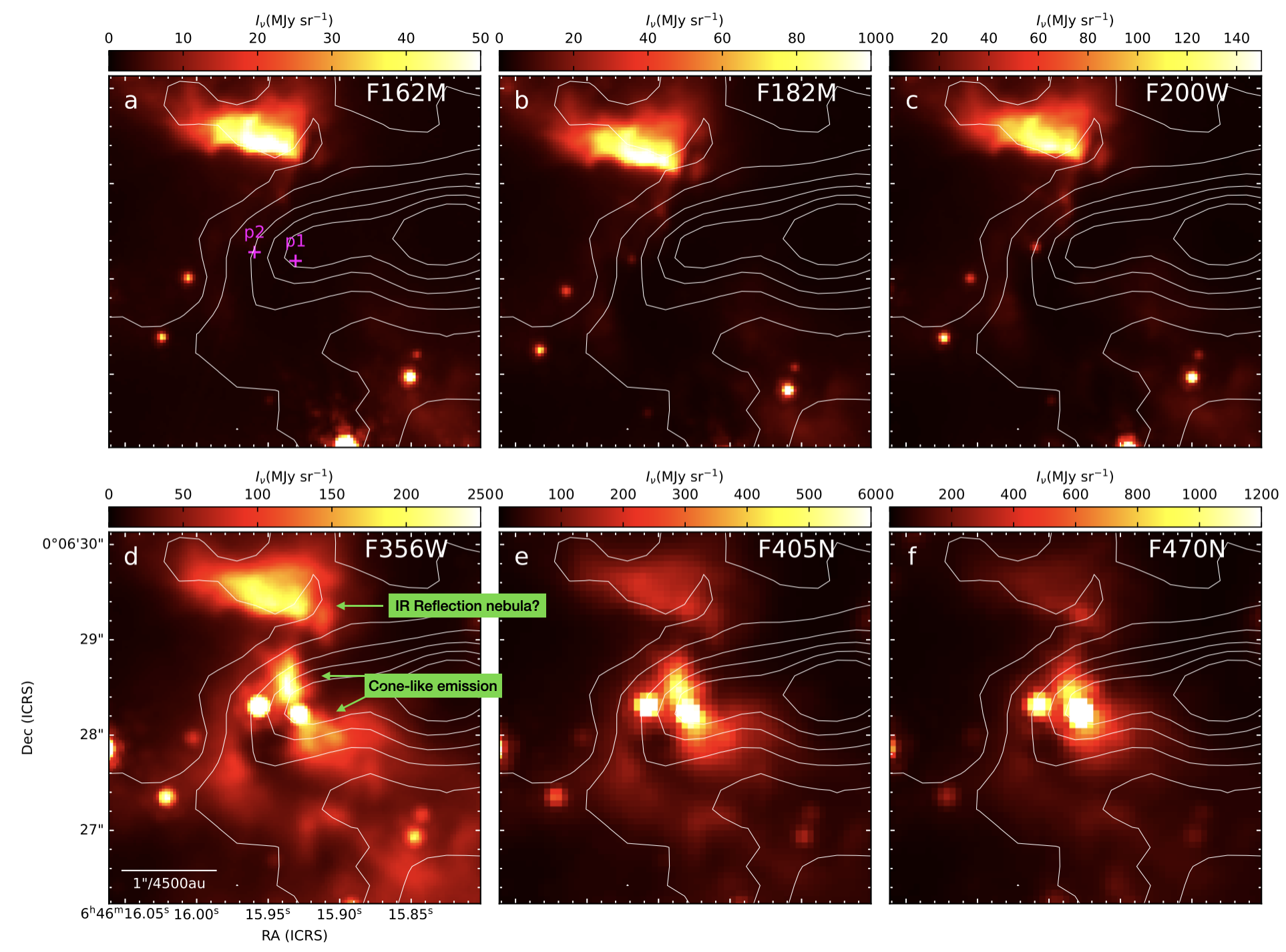}
\caption{A zoom-in view of the near-IR emission surrounding massive protostar S284p1. {\it (a)} NIRCam F162M band image in colorscale overlaid with 1.3~mm continuum in contours. The contours levels are (5, 10, 20, 30, 40, 50) $\times $ 1.1~\jypbm. The magenta crosses show the positions of the protostars S284p1 and S284p2.
{\it (b)}  Same as {\it (a)} but for F182M.
{\it (c)}  Same as {\it (a)} but for F200W.
{\it (d)} Same as {\it (a)} but for F356W.
{\it (e)} Same as {\it (a)} but for F405N.
{\it (f)} Same as {\it (a)} but for F470N.}
\label{fig:zoomin}
\end{figure*} 

The {\it JWST}/NIRCam data further reveal the near-IR (NIR) counterparts of the dense cores (\autoref{fig:overview}d, \autoref{fig:outflow_inner}a, \autoref{fig:zoomin}). While c1b appears to have most of its NIR emission blocked by high extinction, c1a (i.e., S284p1) is clearly visible at 3.56~$\mu$m and longer wavelengths (i.e., in the F356W, F405N and F470N images). 
S284p1 is best resolved in the F356W band, which reveals a point source from the protostellar object surrounded by cone-like emission. Such morphology is expected for an outflow cavity centrally illuminated by a massive protostar. Thus much of the observed F356W continuum emission is mostly likely dominated by scattered light emerging from outflow cavities, similar to morphologies observed around some lower-mass protostars \citep[e.g.,][]{Federman24}. This cone-shaped emission extends approximately 0\farcs{5} ($\sim$2250~au) on both sides of the protostar, with a half-opening angle of $\sim$25\arcdeg. The orientation of the cone is broadly consistent with that of the main outflow axis. The NE outflow cone is significantly brighter than the SW cone, which suggests that NE outflow is orientated towards our line of sight. This is consistent with there being a greater amount of blue-shifted CO(2-1) emission on the NE side than the SW side (see \autoref{fig:outflow_inner}b), although some blue and redshifted emission is present on both sides, indicating a wide opening angle of the outflow (see also \autoref{fig:alma}b). For narrow band F405N and F470N images, significant emission enhancement is clearly visible around the protostar but the cone morphology is less clear. We performed continuum subtraction but did not find clear evidence for line-emitting structures near the protostar (see \autoref{sec:app}).

In all NIRCam bands there exists another illuminated feature about 1\arcsec{} to the north of c1a, which also has a cone-like shape and shows similar morphology as the low level 1.3~mm emission. It is possible that this is an IR reflection nebula that is illuminated by c1a or another source in the region. There is additional IR nebulosity to the south and east of c1a, where several point sources, likely to be YSOs, are also seen. The most prominent of these points sources, which we label S284p2, is approximately 3000~au to the east of S284p1 {(see \autoref{fig:overview}d)}. Given its SED, we consider that S284p2 is likely to be an embedded intermediate-mass YSO based (see \S\ref{sec:sed}). 

\subsubsection{Properties of S284p1 from SED fitting}\label{sec:sed}

\begin{figure*}[ht!]
\includegraphics[width=\textwidth]{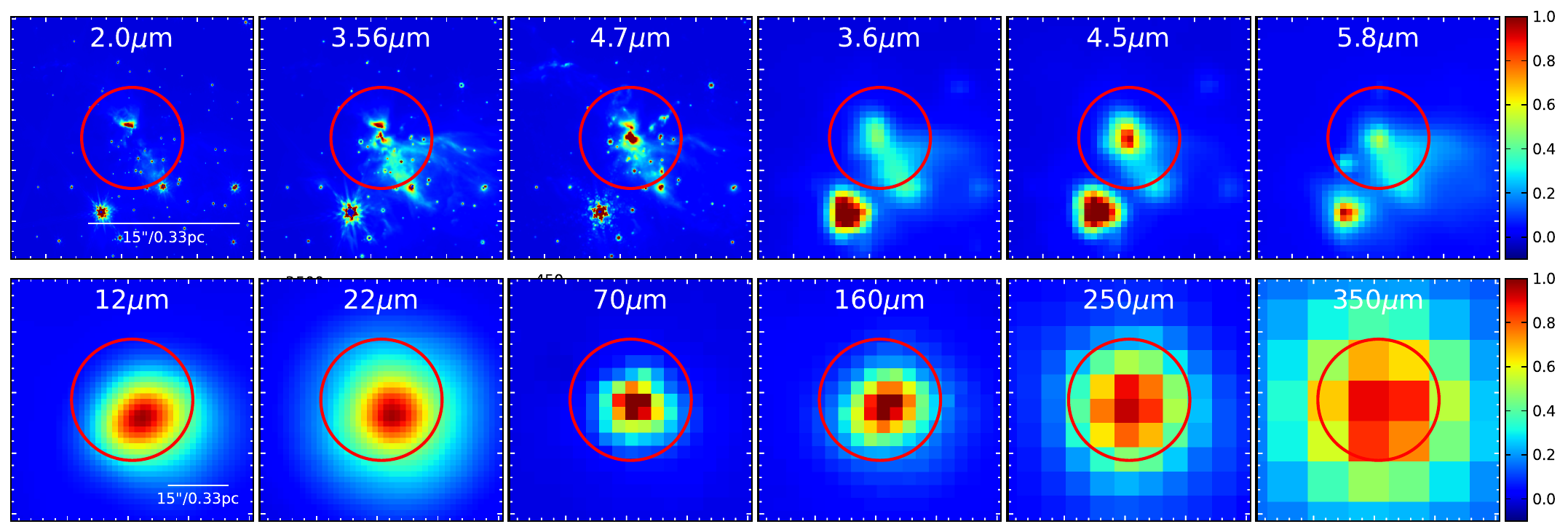}
\caption{Maps of S284c1 in different wavelengths observed with {\it JWST}, {\it Spitzer}, {\it WISE}, and {\it Herschel}. 
The red circle in each panel indicates the aperture used for photometry for the fiducial SED analysis. These apertures have a radius of 15\arcsec\ for wavelengths $>10\:{\rm \mu m}$ and 5\arcsec\ for shorter wavelengths.
}
\label{fig:sed_image}
\end{figure*} 

To determine the properties of the protostar S284p1, we performed a spectral energy distribution (SED) analysis. In addition to the JWST/NIRCAM data, we also retrieved the following archival datasets: {\it Spitzer} 3.5, 4.5, 5.8, and 8.0~$\mu$m; WISE~12, 22~$\mu$m; and {\it Herschel} 70, 160, 250, 350, and 500~$\mu$m (see \autoref{fig:sed_image}). We conduct the SED analysis following the methods of \citet{Fedriani23}, which is based on radiative transfer (RT) models developed within the framework of turbulent core accretion theory \citep[][hereafter ZT18]{Zhang18} and implemented with the python package {\it sedcreator}. This method has been previously applied to Galactic high- and intermediate-mass protostars in a wide range of environments \citep[e.g.,][]{DeBuizer17,Liu19,2020ApJ...904...75L,Fedriani23,Telkamp24}. 

The three primary physical parameters of the protostellar model grid are the initial mass of the core ($M_c$), the mass surface density of the clump environment that the core is embedded in ($\Sigma_{\rm cl}$), and the current protostellar mass ($m_*$), which describes the evolutionary state of a given core. Then the two remaining secondary parameters are the angle of the line of sight to the outflow axis ($\theta_\mathrm{view}$) and the level of foreground extinction ($A_V$). Other model parameters, such as initial core radius ($R_{\rm core}$), remaining core envelope mass ($M_{\rm env}$), and protostellar accretion rate ($\dot{m}_*$) are specified by the three primary parameters.

To account for the low metallicity condition of the region, we have computed a new grid of RT models scaling the opacity per unit gas down by a factor of two compared the ZT18 models. These models, to be presented separately in a future paper (Zhang \& Tan, in prep.), are designed to represent massive protostars forming under half-solar metallicity conditions. Below we investigate the systematic differences that arise in the SED fitting process from using the solar and half-solar metallicity grid.

As input for the SED fitting we perform aperture photometry with a fiducial 15\arcsec{}-radius aperture. This radius was defined with the automated method based on the radial gradient of the background-subtracted enclosed flux at 70~$\mu$m \citep{Fedriani23}. We will consider cases where this fixed aperture is used at all wavelengths, which was the standard method of \citep{Fedriani23}. However, we will also consider a case where at shorter wavelengths (i.e., probed by the NIRCam data, i.e., $<5\:{\rm \mu m}$), a smaller aperture of 5\arcsec{} radius is selected to avoid contamination from other sources.

{We note that a major assumption of the SED fitting method is that a single protostar dominates the flux of the SED at each wavelength. In the case that other, secondary sources contribute significantly, then the derived bolometric luminosity and other related properties, such as current protostellar mass, should be regarded as upper limit constraints on the true source properties. In the case of S284p1, there are a number of factors that suggest the assumption of a single, dominant source is reasonable. First, there is a single, dominant protostellar outflow from the region, as traced by both excited H$_2$ emission and high velocity CO emission. As discussed in \S\ref{sec:S284p1_intro}, the next most significant outflow from the local vicinity appears to be driven from the core c1b, but the extent of this outflow (see Fig.~\ref{fig:outflow_inner}) is only $\sim 4\arcsec$, i.e., at least 10 times smaller than the outflow from S284p1. Nevertheless, below we will consider the impact on our results for S284p1 if c1b contributed 50\% of the total flux.
Next, the morphology of the source at FIR wavelengths near the peak of the SED, i.e., at $70\:{\rm \mu m}$ is spatially concentrated and centered on the location of S284p1. Thus, we do not expect significant contributions to the FIR SED from sources in the more extended cluster. However, as noted, at shorter wavelengths, i.e., $<5\:{\rm \mu m}$, secondary sources can become more important, which then motivates their exclusion via use of a smaller aperture around the main source S284p1.}


In terms of estimating the uncertainties associated with the flux measurements we first present cases that follow the methods of \citet{Fedriani23}. However, in the previous SED fitting analyses, data points at wavelengths $\leq 10$~$\mu$m were treated simply as upper limits, since the RT models do not include PAH emission or thermal emission from transiently heated small grains.
To make better use of the {\it JWST}/NIRCam fluxes, we also explore an option for using the JWST data points as valid measurements, instead of upper limits, but allowing an additional 50\% systematic uncertainty to account for the PAH and small dust grain emission. 

In summary, to investigate the sensitivity of the results to the choice of SED fitting method, we examine three cases:
\begin{enumerate}
\item Case 1: The standard method in \citet{Fedriani23}, i.e., using the {\it Spitzer}, {\it WISE} and {\it Herschel} data with the solar metallicity model grid. Here the Spitzer flux densities are included as upper limits.
\item Case 2: Same as (1) but using the 0.5$Z_\odot$ metallicity model grids.
\item Case 3 (the fiducial case): Same as (2), but including the JWST data. The flux densities in F356W, F405N and F470N bands are are measured with an aperture of 5\arcsec{} {to exclude flux contributions from resolved secondary sources. In addition, these fluxes are now treated as measurements, rather than upper limits}. To account for the uncertainties introduced due to PAH emission and emission from small dust grain, we add in quadrature an extra 50\% systematic uncertainty. The Spitzer data at similar wavelengths are not included in the fit. 
\end{enumerate}


\begin{deluxetable*}{cccccccccccccc}
\tabletypesize{\scriptsize}
\tablecaption{Parameters of the Average and Dispersion of ``Good'' Models \label{table:sed}}
\tablehead{
\colhead{Case} & 
\colhead{\#} & 
\colhead{$M_c$} & 
\colhead{$\Sigma_{\rm cl}$} & 
\colhead{$R_c$} & 
\colhead{$m_*$} & 
\colhead{$\theta_{\rm view}$} & 
\colhead{$A_V$} & 
\colhead{$M_{\rm env}$} & 
\colhead{$\theta_{\rm w,esc}$} & 
\colhead{$\dot{m}_{*}$} & 
\colhead{$L_{\rm bol,iso}$} & 
\colhead{$L_{\rm bol}$} &
\colhead{$t_{\rm age}$} \\
\colhead{} & 
\colhead{} & 
\colhead{($M_\odot$)} & 
\colhead{($\rm g \, cm^{-2}$)} & 
\colhead{(pc)} & 
\colhead{($M_\odot$)} & 
\colhead{(\arcdeg)} & 
\colhead{(mag)} & 
\colhead{($M_\odot$)} & 
\colhead{(\arcdeg)} & 
\colhead{($10^{-5} M_\odot \, \text{yr}^{-1}$)} & 
\colhead{($10^4 L_\odot$)} & 
\colhead{($10^4 L_\odot$)} &
\colhead{($10^5$ yr)}
}
\startdata
S284p1-case1 & \#79 & 63$^{+28}_{-19}$ & 0.21$^{+0.27}_{-0.12}$ & 0.13$^{+0.09}_{-0.05}$ & 12.7$^{+7.0}_{-4.5}$ & 38$\pm$16 & 152$\pm$38 & 25$^{+23}_{-12}$ & 47$\pm$13 & 7.9$^{+5.1}_{-3.1}$ & 3.1$^{+7.3}_{-2.2}$ & 2.4$^{+3.1}_{-1.4}$  & 2.4$^{+1.7}_{-1.0}$ \\
S284p1-case2 & \#148 & 107$^{+35}_{-27}$ & 0.11$^{+0.05}_{-0.04}$ & 0.23$^{+0.08}_{-0.06}$ & 11.7$^{+7.2}_{-4.4}$ & 48$\pm$21 & 147$\pm$32 & 68$^{+35}_{-23}$ & 34$\pm$14 & 6.3$^{+1.6}_{-1.3}$ & 1.7$^{+3.3}_{-1.1}$ & 1.9$^{+2.7}_{-1.1}$  & 3.2$^{+1.6}_{-1.1}$ \\
S284p1-case3 	& \#95 	& 110$^{+25}_{-20}$ 	& 0.10$^{+0.01}_{-0.01}$ 	& 0.24$^{+0.04}_{-0.03}$ 	& 10.2$^{+3.9}_{-2.8}$ 	& 58$\pm$18 	& 129$\pm$21 	& 80$^{+31}_{-23}$ 	& 29$\pm$10 	& 6.0$^{+0.6}_{-0.6}$ 	& 0.9$^{+0.4}_{-0.3}$ 	& 1.5$^{+1.1}_{-0.6}$ & 3.1$^{+0.7}_{-0.6}$ \\ 
\enddata
\tablecomments{
Columns from left to right describe: source and case of SED fitting method; number of ``good'' models that are able to fit the SED; initial core mass ($M_c$); mass surface density of the clump environment ($\Sigma_{\rm cl}$); current protostellar mass ($m_*$); angle of the line of sight to the outflow axis ($\theta_{\rm view}$); amount of foreground extinction ($A_V$); mass of the infall envelope ($M_{\rm env}$); half opening angle of the outflow cavity ($\theta_{\rm w,esc}$); 
accretion rate of the protostar ($\dot{m}_{*}$); bolometric luminosity assuming isotropic emission of the flux ($L_{\rm bol,iso}$); true intrinsic bolometric luminosity ($L_{\rm bol}$); and age of the protostar ($t_{\rm age}$). 
}
\end{deluxetable*}

\begin{figure*}[!ht]
\centering
\includegraphics[width=1\textwidth]{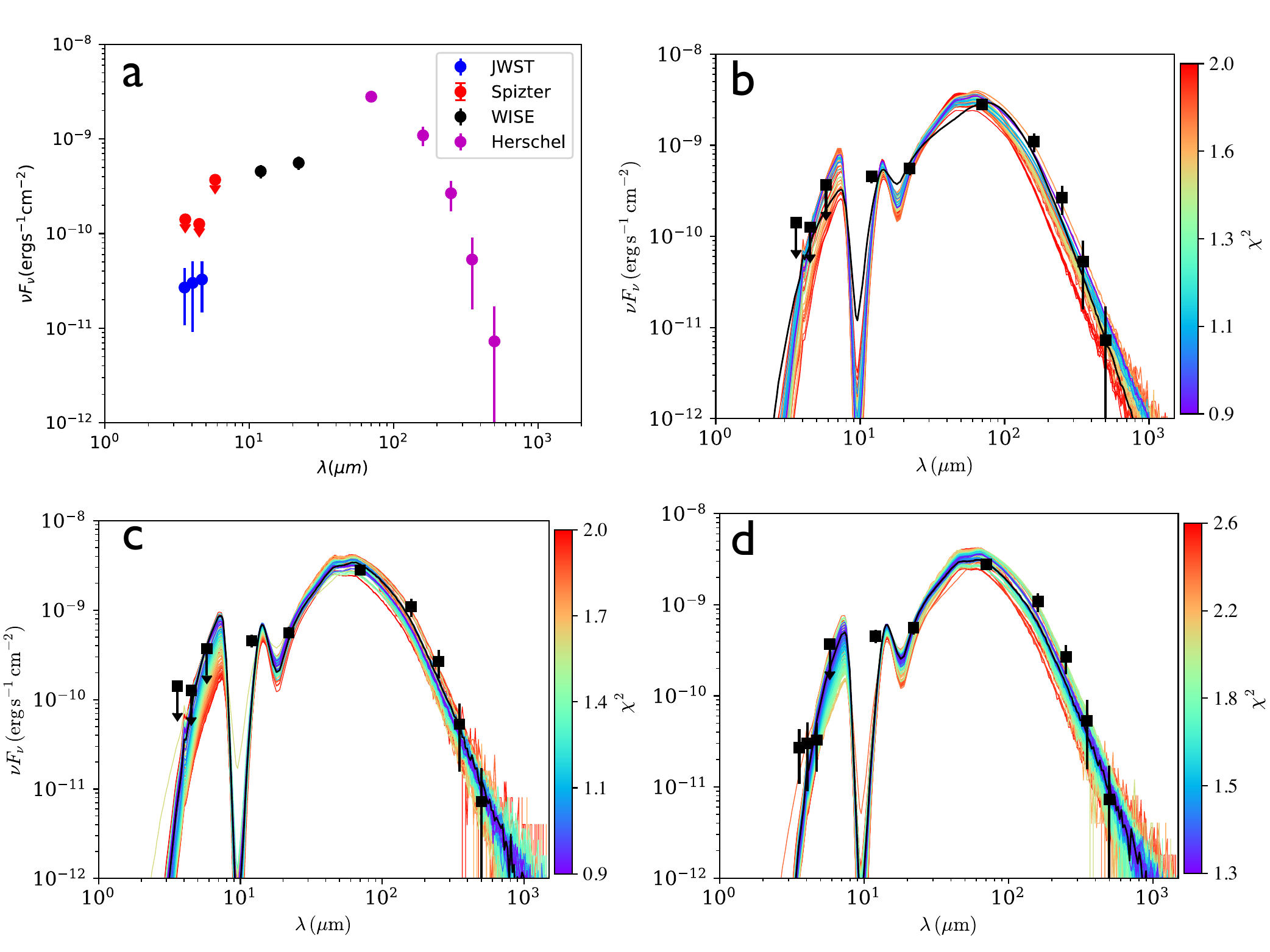}
\caption{ {\it (a)} Protostellar SEDs with fluxes that define the SED for S284p1, with telescope facilities labelled in the legend. {\it (b)-(d)} The SED fitting results for S284p1 with different fitting set ups. The best fitting protostar model is shown with a black line, while all other ``good'' model fits (see text) are shown with colored lines (red to blue with increasing $\chi^2$). {\it (b)} Case 1: fitting uses {\it Spitzer}, WISE and {\it Herschel} data with solar metallicity model grids following method of \citet{Fedriani23}. {\it (c)} Case 2: same as {\it (b)}, but using the half-solar metallicity model grids. {\it (d)} Case 3: same as {\it (c)}, but now fits include JWST/NIRCam data.
}
\label{fig:sed_all}
\end{figure*}

\begin{figure*}[ht!]
\includegraphics[width=\textwidth]{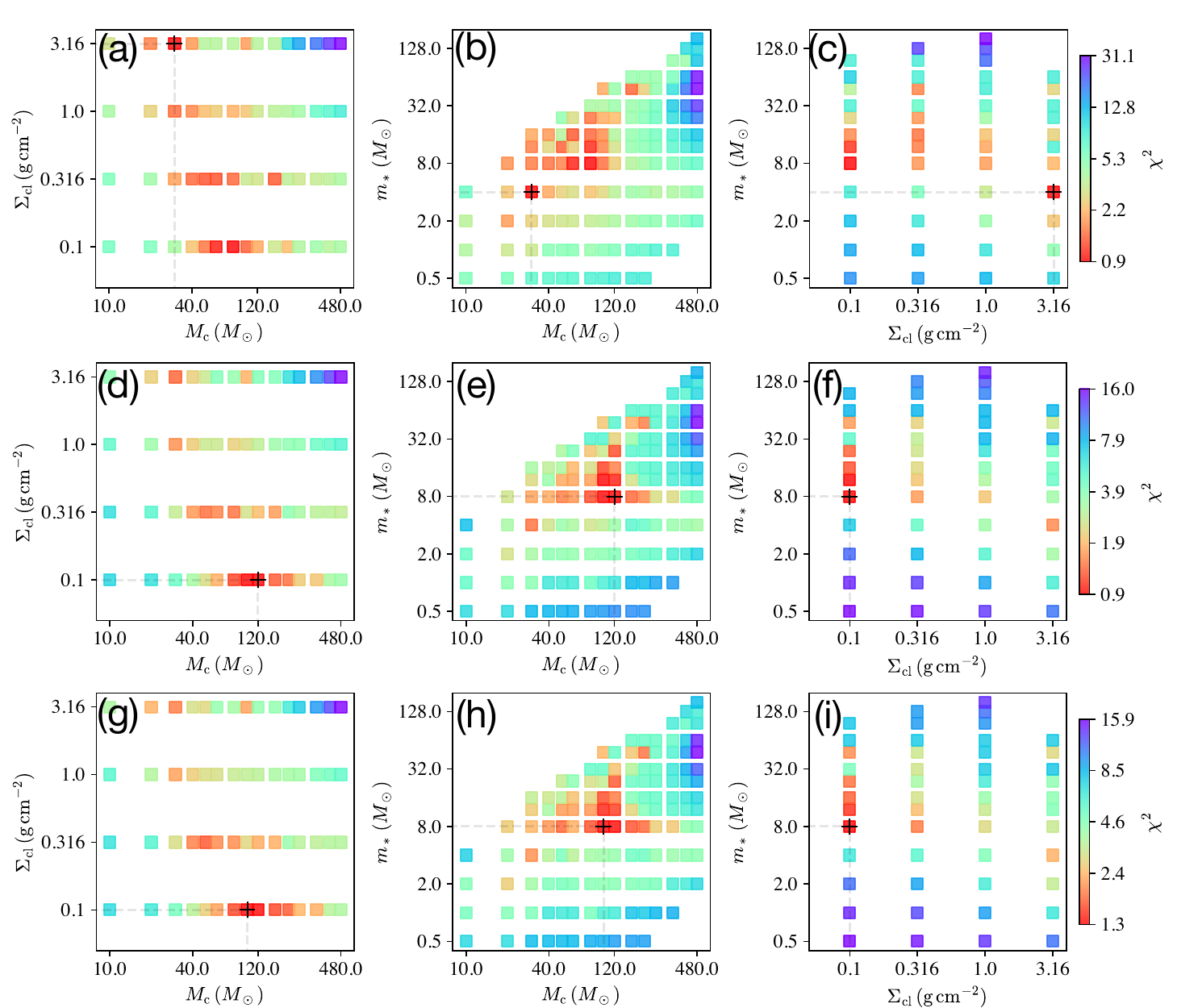}
\caption{Diagrams of the reduced $\chi^2$ distribution of SED model fitting in $\Sigma_{\rm cl}-M_c$ space, $m_\star-M_c$, and $m_\star-\Sigma_{\rm cl}$ space. The black cross and dashed gray lines mark the location of the best model. The white regions are where the $\chi^2$ is larger than 50. Panels (a)-(c), (d)-(f) and (g)-(i) show the SED fit results for cases 1, 2 and 3, respectively (see text). 
}
\label{fig:sed_2d}
\end{figure*} 

In \autoref{fig:sed_all} we show the observed SED of S284p1 (panel a), followed by the SED fitting results of Cases 1, 2 and 3 (panels b, c, d). For the selection of acceptable, i.e., ``good'', models, {we follow the methods of \citet{Telkamp24}. These are models that satisfy $\chi^2<2$ (or the 10 best physical models if $\chi^2_{\rm min}>2$), where $\chi^2_{\rm min}$ is the minimum value of the $\chi^2$ statistic obtained across all models.} The average and dispersion of the parameters of all ``good'' RT models are listed in \autoref{table:sed}.


For the fiducial Case 3, we derive a current protostellar mass of $m_*=10.2^{+3.9}_{-2.8}$~\msun, forming from a core with initial mass of $M_c\sim 110\:M_\odot$ that was embedded in a clump environment with a mass surface density of $\Sigma_{\rm cl}=0.10^{+0.01}_{-0.01}$~\gcm. The protostar is estimated to be accreting at a rate of $\dot{m}_*=6.0^{+0.6}_{-0.6}\times 10^{-5}$~\msunyr, while the current core envelope mass is $M_{\rm env}\sim 80\:M_\odot$. The bolometric luminosity of the source is $\sim 1.5\times 10^4\:L_\odot$ and the age of the system is $t_{\rm age}\sim 3\times 10^5\:$yr.

We see that several of the derived parameters align well with other observational constraints. For example, the estimate of the current envelope mass agrees well with the mass of core c1, which we estimated from the ALMA 1.3~mm flux to have a mass of $68\:M_\odot$ (but based on the uncertain assumption of a dust temperature of 20~K). We note that given the aperture radius of 15\arcsec{} used for SED fitting at longer wavelengths, this mass estimate includes the entire c1 core. Thus, protostellar properties of initial core mass and current envelope mass that are at the lower end of the allowed ranges, i.e., $M_c\sim 90\:M_\odot$ and $M_{\rm env}\sim 60\:M_\odot$, may be more preferred.
Additional properties that are derived include the best-fit clump mass surface density, which matches the estimates from the {\it Herschel} derived column density map (i.e., 0.15~\gcm). 
Finally, the half opening angle of the outflow cavity, $\theta_{\rm w,esc}=29\pm10$\arcdeg, is consistent with the estimate from the IR morphology seen in the JWST/NIRCam images ($\sim$25\arcdeg). 

We see that the impact of using the half-solar metallicity models is to slightly reduce the estimate of current protostellar mass, while more significantly increasing the estimate of initial core mass and current envelope mass. The impact of using the JWST fluxes (compared to the Spitzer-IRAC upper limits) is to tighten the constraints on the received flux, which impacts the isotropic bolometric luminosity, $L_{\rm bol,iso}$. This yields slightly tighter constraints on parameters such as the current protostellar mass, accretion rate and age of the system.

As has been known from previous studies \citep[e.g.,][]{Fedriani23}, SED fitting is prone to significant degeneracies. These are illustrated for the three primary model parameters in \autoref{fig:sed_2d}. However, one of the main conclusions that can be drawn from these distributions is that the inference of a relatively high protostellar mass, i.e., with $m_*\sim 10\:M_\odot$ appears to be relatively robust. However, it should be noted that the models assume a single dominant source is responsible for the observed fluxes. {If another protostar (such as S284p2 or the one in core c1b) contributes to this flux significantly, e.g., at the 50\% level, then the mass estimate for S284p1 would be reduced, but only by a relatively modest amount, i.e., from $m_*=10.2^{+3.9}_{-2.8}\:M_\odot$ to $m_*=9.1^{+2.9}_{-2.2}\:M_\odot$.}

{On the other hand, we also attempt to assess the properties of S284p2 with SED fitting, as the source is clearly resolved in the NIRCam images. Although its flux contribution at longer wavelengths is not well constrained, our analysis suggests that S284p2 can be interpreted as an intermediate-mass protostar viewed through significant extinction (see \autoref{sec:app_p2}).}


\subsubsection{The S284p1 outflow}\label{sec:outflow}

\begin{figure*}[ht!]
\centering
\includegraphics[width=1.\textwidth]{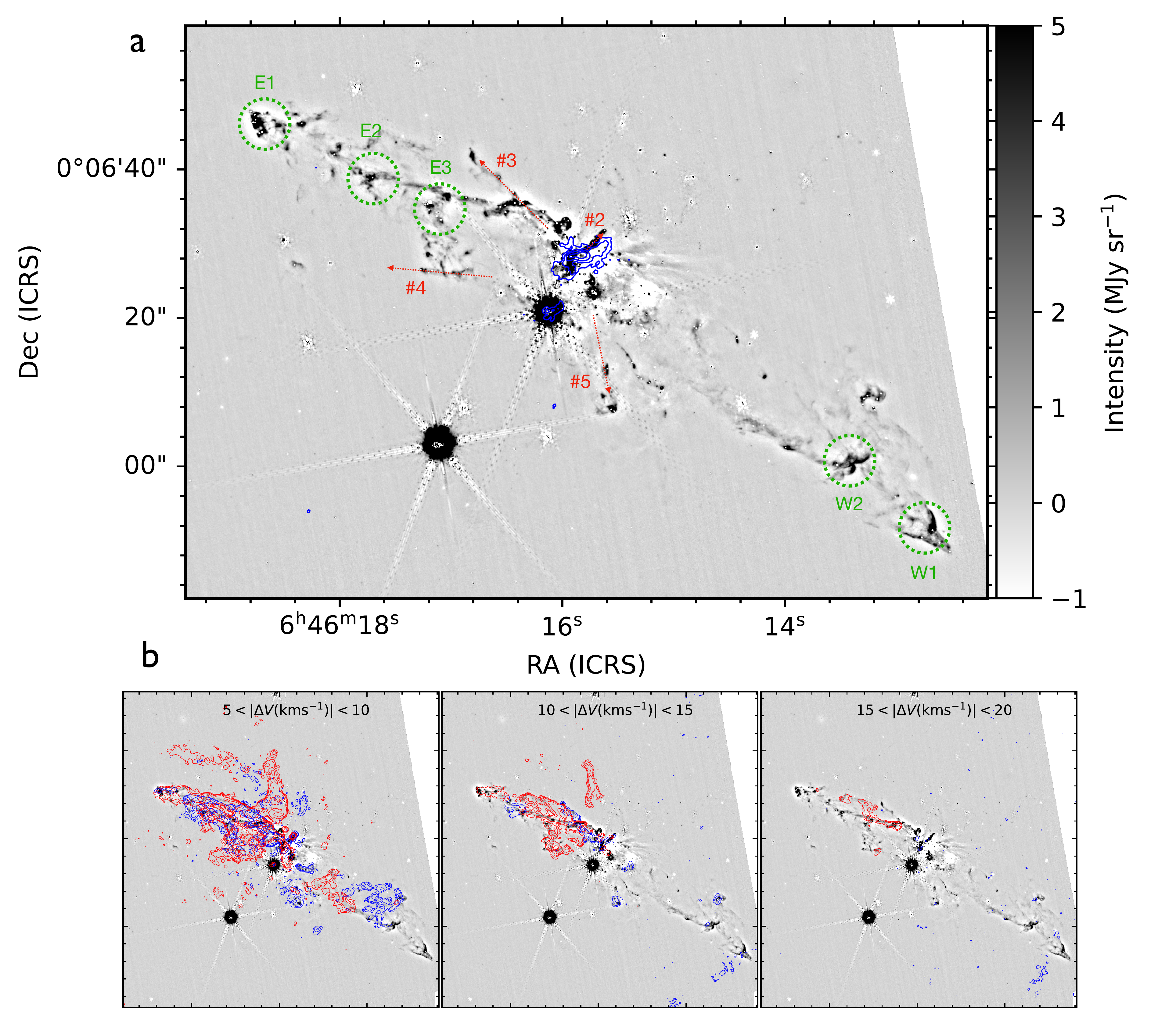}
\caption{The S284p1 massive protostellar outflow seen in the \htwo{} 4.7~$\mu$m{} and CO(2-1) lines. {\it (a)} The F470N image after continuum subtraction. The blue contours show ALMA 1.3~mm continuum emission with contour levels of (5, 10, 20, 40, 80) $\times$ 0.11~\mjypbm. The green circles and red arrows denote the prominent knots in the main jet and other possible jets in the region. {\it (b)} The blueshifted and redshifted CO~(2-1) emission integrated in three velocity ranges from 5 to 20~\kms{} relative to the systemic velocity.}\label{fig:outflow}
\end{figure*} 

{To search for \htwo{} line-emitting structures, we present the continuum-subtracted F470N image in \autoref{fig:outflow}a. The continuum level is estimated by extrapolation from the F356W band flux densities. The idea is that the continuum emission in the F470N is most likely dominated by the nebulosity on larger spatial scales, while shock knots that emit H2 lines are expected to be relatively compact. To isolate such features, we first derive a ratio map for $F$(4.70~$\mu$m)/$F$(3.56~$\mu$m) on a pixel-by-pixel basis and smooth it with a gaussian kernel with a FWHM of 15\arcsec. We then take this smoothed ratio map as a proxy for the nebulosity continuum ratio between two bands. We have tested different FWHM sizes and found that 15\arcsec{} provide optimal compromise between capturing broad-scale nebulosity and minimizing artifacts. We then multiply the F356W image with this ratio map to estimate the continuum contribution in the F470N band, which is subsequently subtracted from the original F470N image.}
We note that the F356W image could contain lines from \htwo{} so the 4.70~$\mu$m line emission could potentially be somewhat over-subtracted. 

From \autoref{fig:outflow}a we see that the global extent of the outflow appears to be contained in the image. The furthest extent of the eastern outflow is defined by knot E1, while that of the western outflow by knot W1. With the central driving source, i.e., S284p1, identified in the JWST/F356W images located at RA, DEC of 
($6^{\mathrm{h}}\,46^{\mathrm{m}}\,15.93^{\mathrm{s}},\ +0^\circ\,6'\,28\farcs{2}$).
From this position we can draw vectors that extend to the furthest extent of the outflows seen in the image (\autoref{fig:overview}b). The morphology of the $\rm H_2$ emission at these extremities suggests that these are collimated bow shocks associated with the outflow propagating into the ambient ISM. The positions defining the end points of the outflow, i.e., locations of E1 and W1, are not as well defined as that of S284p1. However, we set them to be 
($6^{\mathrm{h}}\,46^{\mathrm{m}}\,21.16^{\mathrm{s}},\ +0^\circ\,6'\,47\farcs{09}$)
for the NE outflow and 
($6^{\mathrm{h}}\,46^{\mathrm{m}}\,12.67^{\mathrm{s}},\ +0^\circ\,5'\,51\farcs{11}$)
for the SW outflow. Thus the position angles of the NE and SW vectors are 66.3\arcdeg{} and 232.8\arcdeg, respectively. Their linear extents in the plane of the sky are 46\farcs{5} (1.01~pc) and 61\farcs{2} (1.34~pc), respectively, i.e., an average length of 1.18~pc.

If E1 and W1 represent the location of the outer bowshocks of the outflow, then we can assess the average speed that they have propagated through the interstellar medium given the estimated age of the protostar, i.e., $t_{\rm age}\sim 3\times 10^5\:$yr. In the plane of the sky, this average speed is 3.8~km/s. The true 3D length and average speed of the outflow is greater than the plane of sky component by $({\rm sin\:}\theta_{\rm view})^{-1}$. The value of $\theta_{\rm view}$ is not very well constrained from the SED fitting, but if we adopt a value of $60^\circ$, then the 3D length and average speed are only boosted by a factor of 1.15, i.e., an implied average 3D speed of 4.4 km/s. We note that a near plane-of-sky orientation for the outflow axis is supported by the fact that both blue and redshifted CO emission components are seen in the eastern outflow.

The speed of the primary wind material launched in protostellar outflows is expected to increase as the mass of the protostar grows. Thus it is possible that the current extent of the outflow, as traced by E1 and W1, is driven by more recent outflow activity. For example, if the material defining E1 and W1 was launched $10^5\:$yr ago, then the average propagation speed would be three times greater than the above estimates, i.e., $\sim 13\:$km/s. Furthermore, the current propagation speed, e.g., as may be possible to measure via future proper motion studies, would be expected to be larger than the average.

\autoref{fig:outflow}a reveals other $\rm H_2$ emission features along the outflow axis. On the western side, we identify the W2 knot complex, which, like W1, also exhibits bowshock-like morphology. W2's location is about 75\% of the distance to W1. On the eastern side, after E1, the next most prominent $\rm H_2$ emission feature is the E2 complex, which also shows bow-shock morphology. It is located about 65\% of the distance to E1. It is thus possible that W2 and E2 were caused by a common event of increased level of outflow activity that occurred after the earlier outflow activity that produced W1 and E1. It is possible that such outflow activity has been driven by a relatively transient accretion ``burst''. However, it could also reflect a more sustained level of faster outflow speeds. Such an increase in outflow speed is expected as the protostar grows in mass and/or shrinks in size \citep[e.g.][]{Staff23}. We also note that on the eastern side, there are a greater number of other outflow knots present, including the prominent E3 complex that is interior to E2. As marked in \autoref{fig:outflow}a, there are also several features that may trace outflows driven by other protostars in the region.


Next we consider the outflow traced by CO(2-1) emission. \autoref{fig:outflow}b presents the CO(2-1) integrated intensity map over different velocity ranges relative to the systemic velocity, $v_{\rm sys}=41.2\:$\kms, which is determined from dense gas tracers, such as \ceighteeno(2-1) and DCO$^+$(3-2). 
On the eastern side, where there is a greater level of CO emission than the western side, the CO(2-1) emission exhibits a spatial extent similar to that of the \htwo{} 4.7~$\mu$m line. 
The relative lack of CO emission on the western side may indicate a stronger FUV radiation field in this location, which would be consistent with a scenario of greater exposure to the radiation field from the Dolidze 25 young cluster, which is to the west of S284-FIR1. If the western molecular outflow has suffered a greater degree of photodissociation, then it should be correspondingly brighter in emission from atomic and/or ionized carbon.


Considering kinematics, as shown in \autoref{fig:outflow}b, most CO emission is present at velocity amplitudes from 5 to 10~km/s away from the systemic velocity, with smaller amounts from 10 to 15~km/s and very minor components from 15 to 20~km/s.
The overlapping of blueshifted and redshifted emission in both the eastern and western lobes suggests that the global outflow orientation is close to the plane of the sky. However, if the protostellar outflow has evolved to have relatively wide-angle driving at the core scale, then a large range of both blue and redshifted gas is expected, especially in the inner regions.


We measure the integrated fluxes of the eastern outflow lobe in order to estimate its associated mass. The column densities are derived from the measured line fluxes, using the calcu toolkit \citep{Li20}, for which local thermodynamic equilibrium (LTE) conditions and optically thin line emission have been assumed. We assume a constant excitation temperature of 20~K and a $^{12}$CO abundance of $\rm 0.5 \times 10^{-4}$ with respect to $\htwo$, i.e., half the canonical abundance of solar neighborhood clouds \citep{Blake87}. For the blueshifted component (from 30 to 40.5 km/s) we estimate a mass of $4.6\:M_\odot$ (with innermost 1.5~km/s channel contributing $1.6\:M_\odot$), while for the redshifted component (from 43.5 to 67~km/s) we derive $8.4\:M_\odot$ (with innermost channel contributing $1.8\:M_\odot$). Thus the total mass traced by CO(2-1) emission in the eastern lobe is about $13\:M_\odot$. The total may be underestimated by a few solar masses, since we have excluded the innermost $\pm1\:$km/s about the systemic velocity.

We can compare the above CO-traced mass with the amount of core mass swept up by outflow feedback from the best-fitting SED models (\S\ref{sec:sed}). For case 3, the half opening angle of the outflow cavity is estimated to be $\theta_{\rm w,esc}=32^\circ$. Thus the fraction of one hemisphere of the initial core mass that is swept-up by one outflow cavity is $1-{\rm cos}\:\theta_{\rm w, esc} \rightarrow 0.15$ (assuming a spherical core). For the best fit initial core mass of $110\:M_\odot$, this implies a swept-up mass of $\sim9\:M_\odot$. Although the systematic uncertainties are at a level of at least a factor of two, the approximate agreement between the observed and theoretical mass estimates of the outflow lobe suggests further support for the core accretion model.

While observations of outflow momentum fluxes can, in principle, constrain protostellar models \citep[see, e.g.,][]{2024ApJ...966..117X}, this becomes highly uncertain when the orientation of the system is close to the plane of the sky, i.e., since only a small part of the momentum flux is traced by line of sight motions. Thus, we do not attempt here to measure the momentum flux of the outflow. We note that the maximum observed CO(2-1) velocities from a beam-sized extracted spectrum at S284-p1 find maximum velocities up to about $15\:$km/s from the systemic velocity of the core. Such maximum observed velocities can, {in principle,} be used to help constrain the inclination angles and other properties of core accretion models.

\subsubsection{Implications of the S284p1 outflow geometry for massive star formation theories}\label{sec:outflow_geom}

Here we consider the observed global geometry of the outflow from S284p1 and the implications for massive star formation theories. As discussed, \autoref{fig:overview}b shows annotations that indicate the geometry of the two sides of the outflow. We see that the S284p1 outflow exhibits a high degree of symmetry. The end points of the two sides (E1 and W1 knots shown in \autoref{fig:outflow}a) are well aligned with a common axis, i.e., their angular offset from a common axis is in total only 13.5\arcdeg, or $\lesssim 7^\circ$ for each side of the outflow. Similarly, the total plane of sky projected outflow lengths are very similar, i.e., each deviates from the average length of 1.18~pc by $\lesssim15\%$. Variations in the density of the ambient ISM on either side of S284p1, as well as jet precession, could lead to such differences in outflow directions or extensions. Another optional scenario is that if S284p1 has a plane-of-sky motion in a direction perpendicular to the outflow axis that differs from that of the initially launched gas, then this could lead to the observed level of misalignment. Such motion could result from acceleration acquired by S284p1 during the course of its formation.

To evaluate the level of acceleration that is required, we adopt an estimate of the protostellar age based on SED fitting (\S\ref{sec:sed}) of $t_{\rm age}\sim 3\times10^5\:$yr. If S284p1 started forming at a location that was perfectly aligned with the observed axis connecting the two outflow bowshocks, then it has moved by a projected distance of about 0.14~pc in the plane of the sky. The magnitude of (constant) acceleration needed to achieve this displacement over the course of $3\times 10^5\:$yr is $a\sim 10^{-8}\:{\rm cm\:s}^{-2}$ (leading to current plane of sky motion of about 1 km/s). However, the current observed bowshocks were most likely launched more recently, when the protostar was already relatively massive, i.e., the current observed outflow timescale should be $t_{\rm outflow}<t_{\rm age}$. For example, adopting $t_{\rm outflow}\sim 10^5\:$yr, then requires $a\sim 9 \times 10^{-8}\:{\rm cm\:s}^{-2}$ and a current plane of sky motion of 2.7~km/s. 

The most likely cause of acceleration of S284p1 is gravitational infall towards the core c1b, with the resulting direction of motion being consistent with this scenario. In addition, the expected magnitude of this acceleration is $a_{\rm c1b} = 7\times 10^{-8} (M_{\rm c1b}/50\:M_\odot) (d/0.1\:{\rm pc})^{-2}\:{\rm cm\:s^{-2}}$. Thus, while these estimates are quite uncertain, to order of magnitude the amplitude and direction of acceleration due to gravitational infall towards c1b could explain the apparent offset of S284p1's current location from the outflow axis defined by the pair of outer bowshocks.

Regardless of whether S284p1 has suffered recent acceleration at the above levels, to first order the global outflow morphology is very symmetric. The main conclusion that can be drawn from this symmetry is that the protostar has held a steady orientation over the lifetime of the observed outflow, i.e., especially of its inner disk that is expected to launch the fastest part of the outflow. Such a stable orientation is most naturally expected in core accretion models \citep[e.g.,][]{McKee03} in which the disk is fed by collapse of a relatively symmetric core infall envelope. In contrast, in competitive accretion models \citep[e.g.,][]{2001MNRAS.323..785B,2022MNRAS.512..216G} the accretion (and thus outflow) history is expected to be much more disordered leading to significant changes in outflow orientations. {The observational constraints presented above on the degree of symmetry of oppositely directed outflow bow shocks from a massive protostar motivate the need for quantitative measures of this property as tests of numerical simulations of massive star formation.} We note that the reason why certain numerical simulations of massive star and star cluster formation often show evidence of competitive accretion is likely to be due to initial conditions that do not include regions of dynamically important magnetic fields \citep[see, e.g.,][]{2012ApJ...754....5B,Tan14,2024ApJ...967..157L}.



\section{Discussion}\label{sec:diss}

Massive stars ($\gtrsim$10~\msun) have a major impact on their surroundings via their radiative, mechanical and chemical feedback. Despite this importance, the mechanisms of their formation and how these may vary with environmental conditions remain subjects of intense investigation \citep{Tan14}. Theoretical models predict substantial dependencies on metallicity in the formation, feedback, evolution, and end-products of massive stars \citep[e.g.,][]{Tanaka18}, but direct observations rarely probe down to the immediate surroundings of individual YSOs at a few 1000~au scales due to their typical large distances. The best characterized case to date, HH~1177 in the Large Magellanic Cloud (LMC), is found to be a massive protostar system with a collimated jet and rotating toroid \citep{McLeod18,McLeod24}, which is consistent with the canonical disk-mediated star formation picture. 
However, HH~1177 is relatively isolated. Furthermore, it is an optically revealed source that has evacuated most of its natal material and is thus likely to have evolved beyond its main accretion phase. 

In contrast to HH~1177, S284p1 is deeply embedded in its natal material and actively accreting, thus providing an opportunity to examine low metallicity massive star formation at an the early stage. S284p1 is also seen to be forming in a clustered environment, i.e., surrounded by hundreds of lower-mass YSOs, as well as several other lower-mass protostars driving relatively small jets. In addition, being at a much closer distance, allows resolution of much smaller spatial scales in the case of S284p1 compared to HH~1177.



In this paper, the first of the LZ-STAR survey, we have focused on a detailed, quantitative study of the properties of the most massive protostar in the region, S284p1. We have argued that its highly symmetric outflow morphology is strong evidence for an ``ordered'' formation history, consistent with the simplest expectations of core accretion models. We have developed new massive protostar radiative transfer models designed for the low-metallicity (half-solar) conditions of the S284 region. Fitting these models to the NIR to FIR SED of S284p1 yields estimates of current protostellar mass of about $10\:M_\odot$, with a current age of $\sim3\times 10^5\:$yr. We have explored the degeneracies in protostellar properties that arise from SED fitting. While SED fitting results are quite uncertain, several other pieces of evidence provide quantitative support to the derived model. These include the {\it JWST/NIRCam} observed NIR outflow cavity, which aligns with the global outflow axis and has a half-opening angle that is broadly consistent with that derived from SED fitting. In addition, the amount of CO-traced mass in the outflow lobe is consistent with the material expected to have been ejected from the natal core by the action of outflow feedback. 

We note that some qualitative features of the star formation activity in S284 based on the public {\it ALMA} and {\it JWST} data of the LZ-STAR Survey have been presented by \citet{2025ApJ...980..133J}, however this study, which we have only become aware of in the final stages of the preparation of our paper, is not part of the LZ-STAR Survey and has not influenced our analysis.

For a better understanding of the massive protostar S284p1, higher angular resolution {\it ALMA} observations are needed to better probe its accretion disk, especially to see if its current orientation aligns orthogonally to the outflow axis. Observations of disk kinematics could also yield a dynamical estimate of the current protostellar mass that can be compared to the SED-derived value. {\it JWST} spectroscopic studies of the protostar and its outflow {\citep[see, e.g.,][for an example of such observations]{2023A&A...673A.121B}} should yield improved kinematic constraints on the theoretical models, as well as probing ice absorption features in the infall envelope. Finally, a second epoch NIRCam observation is needed to constrain proper motion kinematics of the outflow features.

\section{Conclusion}\label{sec:sum}


We have undertaken a multi-wavelength survey in the outer Galaxy star formation region Sh2-284 (LZ-STAR Survey). As first paper of the series, we present the ALMA and JWST observations in one of the most active star forming sites in Sh2-284, S284-FIR1. Key conclusions from our study are:

\begin{itemize}
    \item ALMA 1.3~mm continuum observations reveal a population of 22 dense cores with masses ranging from $\sim0.37\,M_\odot$ to $\sim68\,M_\odot$. A subset of these cores drive CO outflows, indicating active ongoing star formation.
    \item The JWST/NIRCam images reveal the presence of a nascent star cluster of about 200 detected members, whose detailed properties are the focus of the next paper in this series. A spectacular bipolar outflow traced by $\rm H_2$(0-0 S(9)) emission is seen to be launched from a protostar near the center of the cluster (in projection). The outflow spans a total length of 2.6~pc on the sky and is also clearly seen in CO(2-1) emission, with these kinematics indicating an orientation near that of the plane of the sky.
    \item ALMA observations reveal that the dense core hosting S284p1 (c1) is fragmented into two sub-cores, with S284p1 residing in a $\sim20\,M_\odot$ component c1a. The JWST/NIRCam F356W image reveals S284p1 as a point source surrounded by a well-defined, cone-shaped near-IR cavity.
    \item We have developed radiative transfer models of massive protostars designed for the low metallicity of S284 and the resulting SED analysis yields a current protostellar mass of $\sim10^{+4}_{-3}\,M_\odot$, which has been forming for last $\sim 3\times10^5\:$yr.
    \item The symmetric, bipolar outflow observed in both CO and \htwo{} emission suggests that S284p1 has undergone a stable, disk-mediated accretion over its formation history that is consistent with basic expectations of core accretion models. 

\end{itemize}


\acknowledgments
This paper makes use of the following ALMA data: ADS/JAO.ALMA\#2017.1.00552.S. ALMA is a partnership of ESO (representing its member states), NSF (USA) and NINS (Japan), together with NRC (Canada), MOST and ASIAA (Taiwan), and KASI (Republic of Korea), in cooperation with the Republic of Chile. The Joint ALMA Observatory is operated by ESO, AUI/NRAO and NAOJ. The National Radio Astronomy Observatory is a facility of the National Science Foundation operated under cooperative agreement by Associated Universities, Inc.
Y.C. was partially supported by a Grant-in-Aid for Scientific Research (KAKENHI  number JP24K17103) of the JSPS.
J.C.T. acknowledges support from NSF grant AST-2206450 and ERC Advanced grant 788829 (MSTAR).
R.F. acknowledges financial support from the Severo Ochoa grant CEX2021-001131-S MICIU/AEI/ 10.13039/501100011033 and PID2023-146295NB-I00. The US-based team members acknowledge support from JWST grant GO-02317.

\vspace{10mm}
\facilities{Atacama Large Millimiter/submillimeter Array (ALMA), Hubble Space Telescope (HST)}
\software{CASA \citep{McMullin07}, APLpy \citep{Robitaille12}, Astropy \citep{Astro13,astropy18,astropy22}}

\clearpage

\appendix
\counterwithin{figure}{section}
\counterwithin{table}{section}


\section{Local continuum subtraction for JWST narrow bands F405N and F470N}\label{sec:app}



\begin{figure*}[ht!]
\centering
\includegraphics[width=0.8\textwidth]{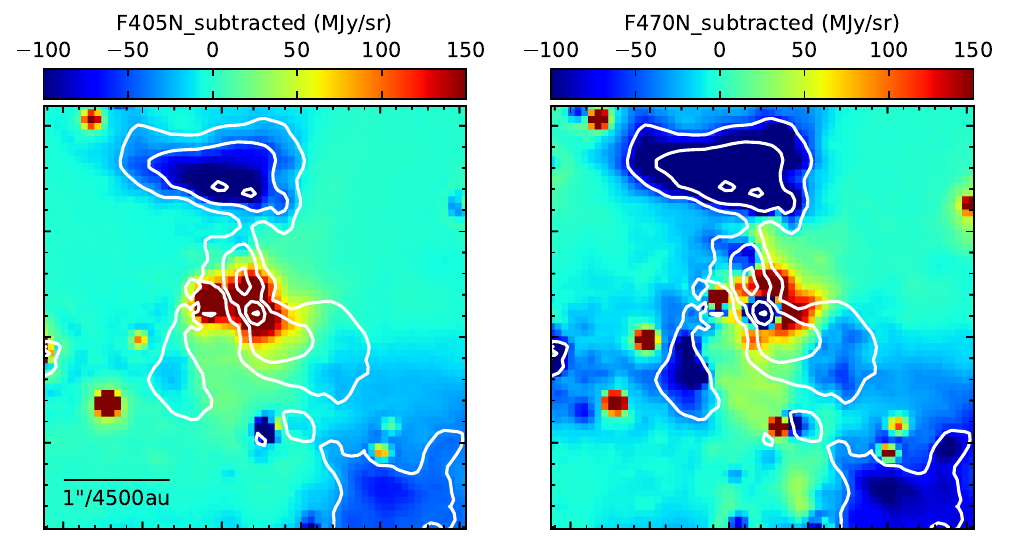}
\caption{Continuum-subtracted image for F405N and F470N for the surrounding of S284p1. {\it (Left)} Continuum-subtracted F405N image. The contours show the F356W image with levels of (50, 100, 200, 400) $\times$ MJy\ sr$^{-1}$. {\it (right)} Same as {\it (Left)} but for F470N. 
}
\label{fig:contsub}
\end{figure*}

In the narrow band F405N and F470N images, a significant emission enhancement is clearly visible around the protostar, although the cone morphology is less distinct. To identify potential line-emitting structures near the protostar (S284p1), we performed continuum subtraction. This is done in a way similar to that described above for F470N on larger scales, i.e.,  the narrow filter images are subtracted by extrapolating the F356W image to estimate the continuum fluxes. Since we focus on the localized region centered on S284p1, we estimate the continuum flux ratio for F(4.05~$\mu$m)/F(3.56~$\mu$m) (or F(4.70~$\mu$m)/F(3.56~$\mu$m)) with that empirically derived for a region within an aperture of 3\arcsec, which is presumably dominated by the diffuse nebular emission. The resulting continuum-subtracted images are shown in \autoref{fig:contsub}. Some residual emission is seen near S284p1, but it is currently unclear if it originates from the excess line emission or potential curvature of the continuum SED.

\section{The SED fit for S284p2} \label{sec:app_p2}

\begin{figure*}[!ht]
\centering
\includegraphics[width=0.5\textwidth]{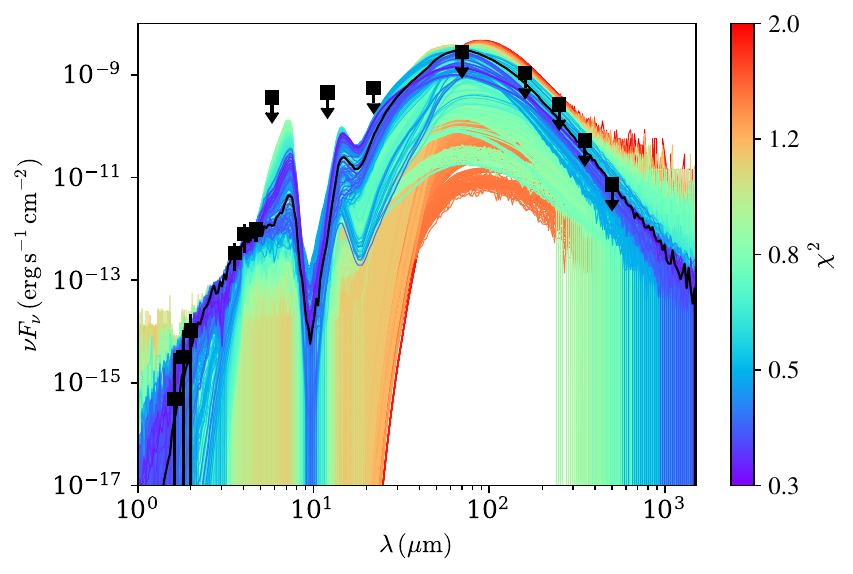}
\caption{SED fitting results for S284p2. The best fitting protostar model is shown with a black line, while all other ``good'' model fits (see text) are shown with colored lines (red to blue with increasing $\chi^2$).
}
\label{fig:sed_second}
\end{figure*}

\begin{deluxetable*}{cccccccccccccc}
\tabletypesize{\scriptsize}
\tablecaption{Parameters of the Average and Dispersion of ``Good'' Models for S284p2\label{table:sed_p2}}
\tablehead{
\colhead{Case} & 
\colhead{\#} & 
\colhead{$M_c$} & 
\colhead{$\Sigma_{\rm cl}$} & 
\colhead{$R_c$} & 
\colhead{$m_*$} & 
\colhead{$\theta_{\rm view}$} & 
\colhead{$A_V$} & 
\colhead{$M_{\rm env}$} & 
\colhead{$\theta_{\rm w,esc}$} & 
\colhead{$\dot{m}_{*}$} & 
\colhead{$L_{\rm bol,iso}$} & 
\colhead{$L_{\rm bol}$} &
\colhead{$t_{\rm age}$} \\
\colhead{} & 
\colhead{} & 
\colhead{($M_\odot$)} & 
\colhead{($\rm g \, cm^{-2}$)} & 
\colhead{(pc)} & 
\colhead{($M_\odot$)} & 
\colhead{(\arcdeg)} & 
\colhead{(mag)} & 
\colhead{($M_\odot$)} & 
\colhead{(\arcdeg)} & 
\colhead{($10^{-5} M_\odot \, \text{yr}^{-1}$)} & 
\colhead{($10^4 L_\odot$)} & 
\colhead{($10^4 L_\odot$)} &
\colhead{($10^5$ yr)}
}
\startdata
S284p2 	& \#5896 	& 75$^{+110}_{-45}$ 	& 0.41$^{+0.97}_{-0.29}$ 	& 0.10$^{+0.14}_{-0.06}$ 	& 4.4$^{+11.6}_{-3.2}$ 	& 58$\pm$21 	& 469$\pm$278 	& 47$^{+114}_{-33}$ 	& 26$\pm$20 	& 9.8$^{+19.8}_{-6.6}$ 	& 0.3$^{+2.6}_{-0.3}$ 	& 0.5$^{+4.2}_{-0.4}$  & 0.8$^{+1.4}_{-0.5}$\\ \hline
\enddata
\tablecomments{
Columns from left to right describe: source and case of SED fitting method; number of ``good'' models that are able to fit the SED; initial core mass ($M_c$); mass surface density of the clump environment ($\Sigma_{\rm cl}$); current protostellar mass ($m_*$); angle of the line of sight to the outflow axis ($\theta_{\rm view}$); amount of foreground extinction ($A_V$); mass of the infall envelope ($M_{\rm env}$); half opening angle of the outflow cavity ($\theta_{\rm w,esc}$); 
accretion rate of the protostar ($\dot{m}_{*}$); bolometric luminosity assuming isotropic emission of the flux ($L_{\rm bol,iso}$); true intrinsic bolometric luminosity ($L_{\rm bol}$); and age of the protostar ($t_{\rm age}$). 
}
\end{deluxetable*}

{S284p2 is resolved in the NIRCam images, so we are able to place some constraints its source properties from SED fitting. \autoref{fig:sed_second} and \autoref{table:sed_p2} present the SED fitting results for S284p2. Located at a distance of approximately $\sim$0\farcs{6} from S284p1, it is challenging to separate the flux densities of the two sources at most wavelengths, except for the JWST bands. Since S284p1 is most likely the dominant source in the region, we adopt the same flux measurements as S284p1 but treat them as upper limits for S284p2. For the JWST bands, where S284p2 is clearly detected, we measure its flux densities using a smaller aperture (1\arcsec) and use these measurements in the SED fitting. The SED fitting results show that a wide range of parameter combinations can reproduce the observed data (owing to the many upper limit measurements). The best-fit solutions suggest that S284p2 can be explained as an intermediate-mass protostar seen through significant extinction. Since it is likely that most of the bolometric flux comes from S284p1, these estimated protostellar properties for S284p2 should mostly be regarded as upper limits, e.g., on current protostellar mass.}

\clearpage


\bibliography{refer}



\end{document}